\documentclass[reprint, aip, footinbib,  letterpaper, superscriptaddress]{revtex4-2}

\usepackage[T1]{fontenc} 

\usepackage{amsmath,color, tikz}
\usetikzlibrary{matrix}
\usepackage{amssymb}
\usepackage{amsfonts}
\usepackage{amsthm}
\usepackage{mathtools}
\usepackage{comment}
\usepackage{hyperref}
\usepackage{cleveref}

\usepackage{algorithm}
\usepackage{algpseudocode}
\usepackage{dsfont}

\usepackage{bbold}
\usepackage{chemformula}
\usepackage{siunitx}

\DeclarePairedDelimiter\bra{\langle}{\rvert}
\DeclarePairedDelimiter\ket{\lvert}{\rangle}
\DeclarePairedDelimiterX\braket[2]{\langle}{\rangle}{#1 \delimsize\vert #2}
\DeclarePairedDelimiterX\braket3[3]{\langle}{\rangle}{#1 \delimsize\vert #2 \delimsize\vert #3}

\usepackage{accents}

\usepackage{fancyvrb}

\usepackage{pdfpages}

\newcommand{\vJ}{\mathbf{J}}
\newcommand{\vP}{{\boldsymbol{\mathcal{P}}}}

\newcommand{\vE}{\mathbf{E}}

\newcommand{\vB}{\mathbf{B}}

\newcommand{\vD}{\mathbf{D}}

\newcommand{\vnabla}{\boldsymbol{\nabla}}

\newcommand{\vH}{\mathbf{H}}

\newcommand{\vR}{\mathbf{r}}

\newcommand{\vxi}{\boldsymbol{\xi}}

\newcommand{\hH}{\hat{H}}

\newcommand{\tr}[1]{\text{Tr}\left(#1\right)}

\newcommand{\red}[1]{{\color{black} #1}}

\makeatletter
\AtBeginDocument{\let\LS@rot\@undefined}
\makeatother

\begin{document}

	\title{Polariton-induced Purcell effects via a reduced semiclassical electrodynamics approach}

        \author{Andres Felipe Bocanegra Vargas}
        \affiliation{Department of Physics and Astronomy, University of Delaware, Newark, Delaware 19716, USA}
	
	\author{Tao E. Li}%
	\email{taoeli@udel.edu}
	\affiliation{Department of Physics and Astronomy, University of Delaware, Newark, Delaware 19716, USA}
	
        \begin{abstract}
        Recent experiments have demonstrated that polariton formation  provides a novel strategy for modifying local molecular processes when a large ensemble of molecules is confined within an optical cavity. Herein, a numerical strategy based on coupled Maxwell--Schr\"odinger equations is examined for simulating local molecular processes in a realistic cavity structure under collective strong coupling. In this approach, only a few molecules, referred to as quantum impurities, are treated quantum mechanically, while the remaining macroscopic molecular layer  and the cavity structure are modeled using dielectric functions. When a single electronic two-level system embedded in a Lorentz medium is confined in a two-dimensional Bragg resonator, our numerical simulations reveal a polariton-induced Purcell effect: the radiative decay rate of the quantum impurity is significantly enhanced by the cavity when the impurity frequency matches the polariton frequency, while the rate can sometimes be greatly suppressed when the impurity is near resonance with the bulk molecules forming strong coupling. Additionally, this approach  demonstrates that the cavity absorption of light exhibits Rabi-splitting-dependent suppression due to the inclusion of a realistic cavity structure. Our simulations also identify a fundamental limitation of this approach --- an inaccurate description of  polariton dephasing rates into dark modes. This arises because the dark-mode degrees of freedom are not explicitly included when most molecules are modeled using simple dielectric functions. As the polariton-induced Purcell effect alters molecular radiative decay differently from the Purcell effect under weak coupling, this polariton-induced effect may facilitate understanding the origin of polariton-modified photochemistry under electronic strong coupling.
	\end{abstract}

	\maketitle

    \section{Introduction}

    Over the past decade, a significant development in chemistry has been the study of molecular polaritons  --- hybrid light--matter states stemming from strong coupling between cavity photon modes and molecular transitions \cite{Hutchison2012,Long2015,Shalabney2015, Dunkelberger2016}.  As summarized in a recent issue of \textit{Chemical Reviews} \cite{Ebbesen2023,Hirai2023,Simpkins2023,Mandal2023ChemRev,Bhuyan2023,Ruggenthaler2023,Tibben2023,Xiang2024}, experimental observations suggest that molecular polaritons offer a novel strategy for modifying various molecular processes, particularly energy transfer \cite{Zhong2017,Xiang2020Science,Chen2022} and chemical reactions \cite{Hutchison2012,Thomas2016,Thomas2019_science,Ahn2023Science}. In this emerging field of polariton chemistry, early experiments predominately involved condensed-phase molecules confined in planar Fabry--P\'erot cavities, studied under both electronic strong coupling (ESC) \cite{Hutchison2012} and vibrational strong coupling (VSC) \cite{Shalabney2015,Long2015} regimes. Recently, the scope of molecular strong coupling experiments has expanded to encompass a growing variety of cavity structures \cite{Chikkaraddy2016,Yoo2021,Brawley2021, Wright2023, Canales2021,Canales2024}. Beyond the study of elementary molecular processes under strong coupling, polariton formation has also been utilized to experimentally design stable blue phosphorescent organic light-emitting diodes (PHOLEDs)  with the polariton-enhanced Purcell effect \cite{Zhao2024}. In this work, plasmon–exciton polaritons are employed to enhance the Purcell factor, or the cavity-induced effects on radiative decay rates of blue PHOLEDs. This enhancement reduces  destructive high-energy triplet  annihilation events, thereby improving the device lifespan.

    Despite tremendous progress in the field of polariton chemistry, recent reports have debated on the proper interpretation of some polariton experiments. For example, can the formation of collective VSC  enhance the Raman scattering of molecules in the absence of external pumping \cite{Shalabney2015Raman}, or is  the enhanced signal merely a result of surface-enhanced Raman scattering  due to the use of gold mirrors \cite{DelPino2015,Ahn2021,Nagarajan2021,Takele2020}? Can the peak shift of highly off-resonant cavity modes serve as an independent indicator of changes in molecular concentration during chemical reactions \cite{Thomas2016}, or is this peak shift instead influenced by changes in Rabi splitting during the reaction \cite{Thomas2024shifts}? Can polariton formation suppress photochemical reaction rates \cite{Hutchison2012}, or does this suppression arise simply  from  reduced cavity absorption of light under strong coupling \cite{Thomas2024}? Can imperfections in cavity mirror placement alter molecular absorption signals under strong coupling \cite{Michon2024}? Overall, these debates underscore that, while it is exciting to attribute various experimental findings to strong coupling, at least some observations may be more straightforwardly explained by more trivial cavity effects.

    Currently, these debates are primarily driven by experimental findings,  promoting the question: can theory and realistic simulations provide valuable insights for addressing these problems? Within the current theoretical toolbox for molecular polaritons \cite{Galego2019,Campos-Gonzalez-Angulo2019,Hoffmann2020,Mandal2020Polarized,Fischer2021,YangCao2021,Wang2022JPCL,DuDark2021,Perez-Sanchez2023,Weight2024,Aroeira2023,Flick2017,Haugland2020,Riso2022,Schafer2021,Bonini2021,Yang2021,McTague2022,Liebenthal2022,Luk2017,Li2020Water,Gomez2023,Lindoy2024,Hoffmann2018,Rosenzweig2022,SangiogoGil2024,Sukharev2023,Sukharev2023a,Zhou2024,Kuisma2022},  a common approach is to describe cavity photons as predefined normal modes. In this approach, very often only a single cavity mode is considered in calculations, a simplification known as the single-mode approximation. While some early theoretical attempts on strong coupling focused on a single molecule coupled to a single cavity mode,  recent research efforts have been made to generalize the single-molecule strong coupling calculations to the collective regime.\cite{Groenhof2019,Li2021Collective,DuDark2021,Perez-Sanchez2023,Lindoy2024,Chng2024JPCL}  For instance, calculations have been performed by explicitly including an increased amount of molecules in the cavity while maintaining the fixed Rabi splitting. \cite{Groenhof2019,Li2021Collective} The asymptotic behavior when the molecular number $N$ becomes macroscopically large may represent the experimental limit.  Alternatively, an effective single-molecule strong coupling model has been developed to describe collective strong coupling by leveraging the permutation symmetry of many molecules. \cite{Perez-Sanchez2023} 
    
    Although significant research efforts have been made on studying collective molecular dynamics coupled to predefined cavity normal modes, many realistic features of cavity geometries remain unaddressed in these studies. For instance,  predefined cavity normal modes may fail to adequately account for the material properties and imperfections of the cavity mirrors, the frequency-dependent cavity absorption and transmission, and the interplay between cavity modes at widely different frequencies (as highly off-resonant cavity modes are typically excluded in normal-mode calculations). As a result, normal-mode-based theoretical approaches may not provide an optimal framework for distinguishing between trivial cavity effects and non-trivial strong coupling effects. 

    To better account for  realistic cavity structures while maintaining  computational efficiency, here, we examine a computational strategy at the interface between the coupled Maxwell--Schr\"odinger equations \cite{Castin1995,Mukamel1999,Lopata2009-1,Lopata2009-2,Deinega2014,Schelew2017,Yamada2018,Li2018Spontaneous,Li2018Tradeoff,Sukharev2017,Zhou2024} and single-molecule strong coupling models. In this computational approach,  only a few molecules --- referred to as  quantum impurities --- are propagated explicitly using  quantum mechanics, while the majority of molecules and the cavity geometry are both described on the level of dielectric functions. With a self-consistent propagation of the coupled Maxwell--Schr\"odinger equations, this approach potentially provides an advantageous solution for describing local molecular processes under collective strong coupling. 
    
    While this semiclassical electrodynamics approach has not been widely used to describe local molecular processes under collective strong coupling, it is an established method for studying the response of a small molecule near plasmonic systems. \cite{Lopata2009-2,Sukharev2017} When modeling collective strong coupling, because most molecules are represented by dielectric functions, this approach significantly reduces the computational cost compared to simulating all confined molecules quantum-mechanically, such as a recent study by Sukharev, Subotnik, and Nitzan on simulating polariton-induced photochemistry in planar Fabry--P\'erot cavities \cite{Sukharev2023,Sukharev2023a}.  Furthermore, since solving Maxwell's equations alone is well-established in the field of computational electrodynamics, we implement this self-consistent algorithm on top of  an existing open-source time-domain solver of Maxwell's equations --- the finite-difference time-domain (FDTD) method \cite{Taflove2005} in the MIT Electromagnetic Equation Propagation (MEEP) package \cite{Oskooi2010}. With this  strategy, the implementation efforts can be  greatly reduced, and this code can be easily adapted to simulate a wide range of cavity geometries. Since our approach does not treat all confined molecules on the same footing (even though the quantum impurities may be identical to the remaining molecules modeled by dielectric functions), we henceforth refer to this method as the \textit{reduced semiclassical electrodynamics approach}. 
    
    With the explicit simulation of only a single or a few quantum impurities, this reduced semiclassical electrodynamics approach  provides a complementary solution for seamlessly adapting single-molecule simulation methodologies from theoretical chemistry \cite{Tully1990,Ben-Nun2000,Li2005Eh} to describe collective strong coupling. Furthermore, due to an explicit consideration of realistic cavity structures, this approach may also aid in distinguishing between trivial cavity effects and non-trivial strong coupling effects, as debated in recent experiments. \cite{Shalabney2015Raman,DelPino2015,Ahn2021,Nagarajan2021,Takele2020,Thomas2016,Thomas2024shifts,Hutchison2012,Thomas2024,Michon2024} To fully realize this potential, however, we need to understand both the advantages and the limitations  of this approach. Specifically, what are the benefits of incorporating realistic cavity structures, and what are the limitations of treating most molecules forming strong coupling as dielectric functions?

    To evaluate the advantages and limitations of the reduced semiclassical electrodynamics approach, in this manuscript, we study a fundamental problem under collective strong coupling: how the radiative decay rate of a single quantum impurity is modified within a realistic cavity structure. To reduce the computation complexity, we model the quantum impurity as an electronic two-level system (TLS). While a large volume of theoretical work in polariton chemistry relies on the single-mode approximation, we simulate a realistic cavity structure containing a well-defined single cavity mode --- a two-dimensional (2D) Bragg resonator composed of periodic dielectric layers. \cite{Skolnick1998,Novotny2006,Kim2007} \red{Inside the Bragg resonator, strong coupling is formed between the confined dielectric medium with the cavity mode, while the  TLS is only weakly coupled to the cavity mode.} 
    
    Using this simplified and well-defined system, we observe a polariton-induced Purcell effect: The radiative decay rate of the impurity is enhanced when the impurity frequency matches the polariton frequency, while the rate can sometimes also be strongly suppressed when the impurity is near resonance with the bulk molecules  forming strong coupling.  Additionally, we discuss the advantages and limitations of this approach as well as the connection between our findings and experiments. 

        \section{Methods}\label{sec:methods}
    Within the reduced semiclassical electrodynamics approach, the quantum impurity is evolved using the following quantum von Neumann equation:
    \begin{subequations}
    \begin{equation}\label{eq:von_Neumann_eq}
        \frac{\mathrm{d}}{\mathrm{d}t}\hat{\rho}(t)=-\frac{i}{\hbar}[\hat{H}_{\rm{sc}}(t),\hat{\rho}(t)]. 
    \end{equation}
    Here, $\hat{\rho}$ denotes the density matrix of the quantum impurity; $\hat{H}_{\rm{sc}}$ represents the semiclassical Hamiltonian which includes the light-matter interaction. If the quantum impurity is initialized in a pure state at $t=0$, i.e., $\hat{\rho}(t=0) = \ket{\Psi(t=0)}\bra{\Psi(t=0)}$, Eq. \eqref{eq:von_Neumann_eq} reduces  to the quantum Schr\"odinger equation, $\frac{\mathrm{d}}{\mathrm{d}t} \ket{\Psi(t)} = \hat{H}_{\rm{sc}}(t) \ket{\Psi(t)}$, where $\ket{\Psi(t)}$ represents the time-dependent wave function of the quantum impurity.  The semiclassical Hamiltonian $\hat{H}_{\rm{sc}}(t)$ is expressed as follows: \cite{Mukamel1999,Li2018Tradeoff,Cohen-Tannoudji1997}
    \begin{align}\label{eq:SCHamiltonian}
    \hat{H}_{\rm{sc}}= \hat{H}_{\rm{s}} -\int d\mathbf{r}\; \mathbf{E}_{\perp}(\mathbf{r},t)\cdot\hat{\vP}(\mathbf{r}), 
    \end{align}
    where $\mathbf{E}_{\perp}(\mathbf{r},t)$ denotes the transverse classical electric field vector, and $\hat{H}_{\rm{s}}$ and $\hat{\vP}(\mathbf{r})$ denote the bare  quantum Hamiltonian and the polarization density operator of the impurity system, respectively. In practice, because   $\hat{\vP}(\mathbf{r})$ is expressed as a spatial Gaussian distribution [see Eq. \eqref{eq:polarization_2d_method} below], the contribution of the longitudinal interaction, $-\int d\mathbf{r}\; \mathbf{E}_{\parallel}(\mathbf{r},t)\cdot\hat{\vP}(\mathbf{r})$, is negligible. Consequently,   the light-matter coupling in Eq. \eqref{eq:SCHamiltonian} can be approximately evaluated using $\int d\mathbf{r}\; \mathbf{E}(\mathbf{r},t)\cdot\hat{\vP}(\mathbf{r})$. \cite{Li2018Tradeoff} This approximation eliminates the need to explicitly evaluate the transverse component and also simplifies the implementation.
    
    The classical electromagnetic field is propagated using the Maxwell's equations in the time-domain: \cite{Taflove2005,Li2018Tradeoff,Li2019Stimulated}
\begin{align}
\frac{\partial}{\partial t}\mathbf{D}(\mathbf{r},t) & =\vnabla \times \mathbf{H}(\mathbf{r},t) - \vJ_{\perp}(\vR, t) , \label{eq:Maxwell1} \\
\mathbf{D}(\mathbf{r}, \omega) &= \epsilon_0\epsilon_{\rm{r}}^{\ast}(\mathbf{r},\omega)\mathbf{E}(\mathbf{r}, \omega) , \label{eq:Maxwell2}\\
    \frac{\partial}{\partial t}\mathbf{H}(\mathbf{r},t) &=- \frac{1}{\mu_0}  \vnabla \times \mathbf{E}(\mathbf{r},t) . \label{eq:Maxwell3}
\end{align}
Here, $\mathbf{D}$, $\mathbf{H}$, $\mathbf{E}$, and $\vJ_{\perp}$ represent the classical displacement field, auxiliary magnetic field, electric field, and transverse current density, respectively. The dielectric medium is assumed to be non-magnetic, so $\vH$ is related to the magnetic field $\vB$ via a simple linear relation: $\vH \equiv \vB/\mu_0$, where $\mu_0$ denotes the vacuum magnetic permeability. $\varepsilon_{0}$ denotes the vacuum permittivity, and the effects of the classical dielectric medium are captured by the relative permittivity $\epsilon_{\rm{r}}^{\ast}(\mathbf{r},\omega)$. 
Eq. \eqref{eq:Maxwell2} is expressed in the frequency domain, meaning that a convolution in the time domain must be evaluated to update $\vE(\vR, t)$ from $\vD(\vR, t)$.  The classical current density $\vJ(\vR, t)$ is generated by the quantum impurity and can be calculated in a mean-field manner as:
\begin{equation}
        \vJ(\mathbf{r},t)= \frac{\mathrm{d}}{\mathrm{d}t}\mathrm{Tr}(\hat{\rho}(t)\hat{\vP}(\mathbf{r})) .\label{eq:current_density_meanfield}
\end{equation}
Its transverse component, $\vJ_{\perp}(\mathbf{r},t)$, can be evaluated by $\vJ_{\perp}(\mathbf{r},t)= \frac{\mathrm{d}}{\mathrm{d}t}\mathrm{Tr}(\hat{\rho}(t)\hat{\vP}_{\perp}(\mathbf{r})) $. In practice, because $\hat{\vP}(\mathbf{r})$ follows a spatial Gaussian distribution, $\vJ(\vR, t)$ is used directly instead of $\vJ_{\perp}(\vR, t)$ when propagating Eq. \eqref{eq:Maxwell1}. It is important to note that if $\hat{\vP}(\mathbf{r})$ were implemented as a point dipole, neglecting the transverse symbols in Eq. \eqref{eq:SCHamiltonian} and Eq. \eqref{eq:Maxwell1} would  result in strong electromagnetic (EM) fluctuations at the dipole location, leading to an incorrect spontaneous emission behavior. \cite{Lopata2009-1,Lopata2009-2}
 \end{subequations}

For simplicity, we model the quantum impurity as an electronic TLS. The quantum Hamiltonian for the TLS is expressed as
\begin{align}\label{eq:LM_Hamiltonian}
    \hat{H}_{\rm{s}}^{(\mathrm{TLS})}=\begin{pmatrix}
0 & 0 \\
0 & \hbar \omega_{\rm{TLS}}
\end{pmatrix}
\end{align}
in the basis $\{\ket{g},\ket{e}\}$, corresponding to the electronic ground and excited states, respectively. Here, $\omega_{\rm{TLS}}$ represents the transition frequency of the TLS. Under the same basis, the polarization density operator is expressed as:
\begin{subequations}\label{eq:polarization_2d_method}
\begin{align}\label{eq:polarization_2d_method-1}
    \hat{\vP}(\mathbf{r})=\vxi(\mathbf{r}-\mathbf{r}_0^{(\mathrm{TLS})})\begin{pmatrix}
0 & 1 \\
1 & 0
\end{pmatrix} ,
\end{align}
where $\mathbf{r}_0^{(\mathrm{TLS})}$ represents the location of the TLS. \red{Substituting Eq. \eqref{eq:polarization_2d_method-1} into Eq. \eqref{eq:current_density_meanfield}, we obtain an analytical form of the current density $\vJ(\vR, t) = -2\omega_{\rm{TLS}}\text{Im}[\rho_{\rm{ge}}(t)]\vxi(\mathbf{r}-\mathbf{r}_0^{(\mathrm{TLS})})$ for numerical calculations, where  $\text{Im}[\cdots]$ denotes the imaginary component.}

Assuming the ground and excited states as the $s$- and $p$-orbitals of a hydrogen atom, the spatial polarization distribution is given by $\vxi(\mathbf{r})=\psi_{\rm{g}}^{*}q\mathbf{r}\psi_{\rm{e}}=(2\pi)^{-3/2}\sigma^{-5}\mu_{12}\mathbf{r}z\mathrm{exp}[-r^{2}/(2\sigma^{2})]$, where $q$ denotes the effective charge, $\sigma$ corresponds to the TLS width, and $\mu_{12}=|\int d\mathbf{r}\psi_{\rm{g}}^{*}q\mathbf{r}\psi_{\rm{e}}|$ quantifies the magnitude of the transition dipole moment between the ground state $(\psi_{\rm{g}})$ and the excited state $(\psi_{\rm{e}})$. \cite{Li2018Spontaneous} Because we restrict ourselves to 2D simulations along the $x$-$y$ plane, the polarization density is expressed as $\vxi_{\rm{2D}}(x,y)= \int_{-\infty}^{+\infty} dz \vxi(\vR)$, leading to 
\begin{equation}
    \vxi_{\rm{2D}}(x,y)= \frac{\mu_{12}}{2\pi \sigma^2}
e^{-\frac{x^2 + y^2}{2\sigma^{2}}} \mathbf{e}_z ,
\end{equation}
where $\mathbf{e}_z$ denotes the unit vector along the $z$-direction. 
\end{subequations}

Our code is implemented on top of the open-source MEEP package, \cite{Oskooi2010} and a comprehensive description of simulation details is provided in Appendix \ref{sec:simu_details}.
    
    \section{Results}

        \begin{figure*}
	    \centering
	    \includegraphics[width=0.85\linewidth]{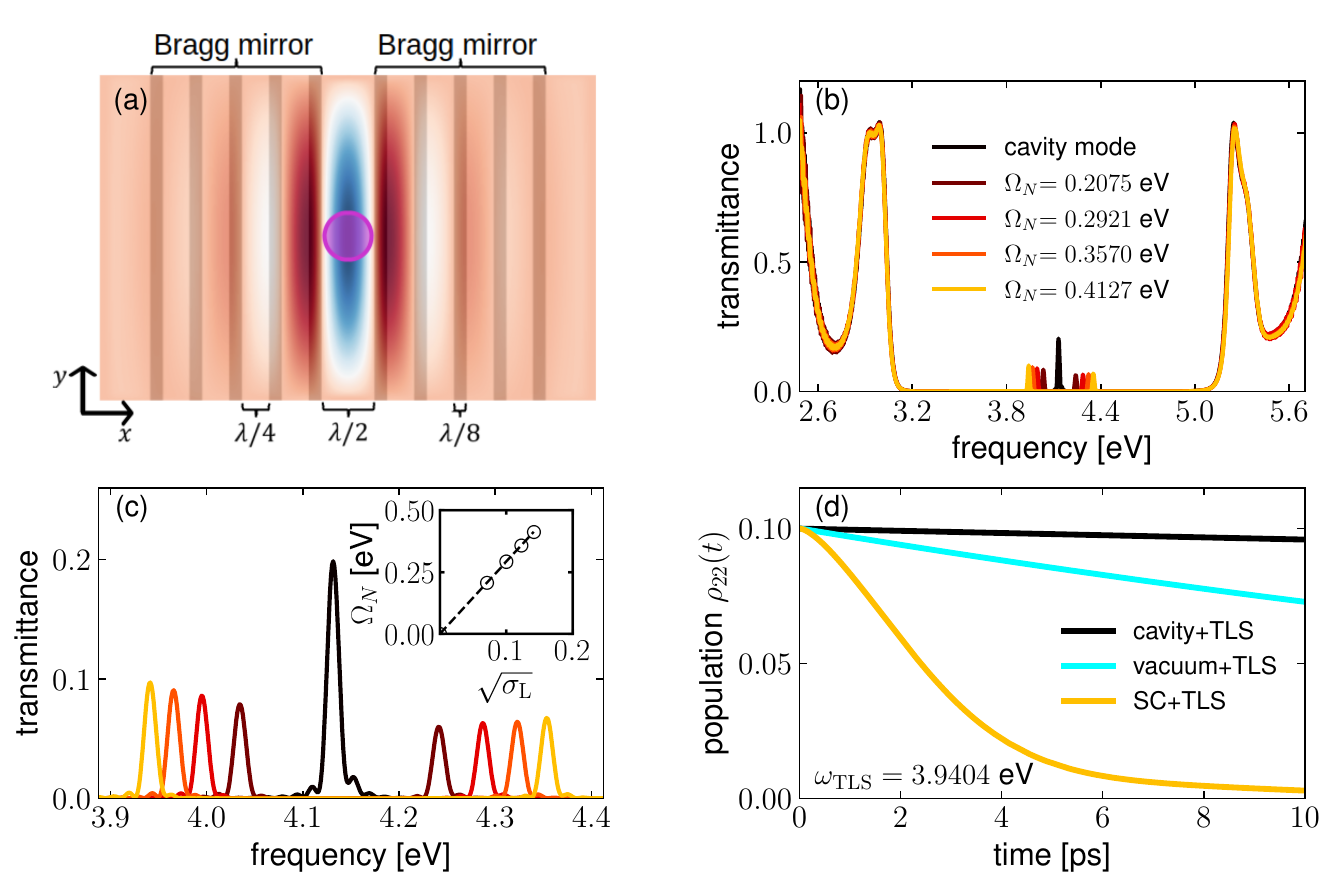}
	    \caption{Simulation setup and excited-state TLS dynamics in the Bragg resonator. (a) 2D cavity geometry formed by a pair of parallel Bragg mirrors, each composed of five layers of equidistant dielectric media (gray rectangles) separated by air. This geometry supports a high-$Q$ cavity mode at $\omega_{\rm{c}}=4.1357$ eV. A Lorentz medium with frequency $\omega_{\rm{L}}=\omega_{\rm{c}}$ can be confined at the center of the cavity to form strong coupling. A snapshot of the electric field distribution (color-coded from red to blue) is included to demonstrate the field confinement inside the cavity. (b) Transmission spectrum of the cavity when the absorption strength of the Lorentz medium is set as $\sigma_{\rm L}=0$ (black, empty cavity), 0.005 (brown), 0.01 (dark red), 0.015 (red), and 0.02 (orange), respectively. The corresponding Rabi splitting values are provided in the legend. (c) Zoom-in polariton spectra corresponding to part (b). The inset plots the linear dependence of the Rabi splitting on $\sqrt{\sigma_{\rm L}}$. (d)  Excited-state population dynamics, $\rho_{22}(t)$, of the TLS with frequency $\omega_{\rm{TLS}}=3.9404$ eV under different conditions: in the empty cavity (black), in vacuum (cyan), and in the strong coupling system (orange). For the case of strong coupling, the Rabi splitting is $\Omega_{N}=0.4127$ eV and the LP frequency matches $\omega_{\rm{TLS}}$. The TLS spontaneous emission is enhanced when its frequency matches the LP frequency.    }
	    \label{fig:setup}
    \end{figure*}

    Fig. \ref{fig:setup}a illustrates the 2D cavity geometry used in our reduced semiclassical electrodynamics simulation\red{; see also Fig. \ref{fig:SI_SetUp} for the detailed description of the simulation setup and system dimensions.} In this cavity setup, a pair of distributed Bragg reflectors (DBRs), or Bragg mirrors, is positioned along the $y$ axis with a separation of $\lambda/2 = 149.89$ nm. Each mirror consists of five layers of equidistant dielectric media (refractive index $n=2$, gray). Adjacent dielectric layers are separated by air  (refractive index $n=1$) with a spacing of $\lambda/4$. Due to the destructive interference of light propagation through the periodic dielectric layers, each Bragg mirror prohibits the transmission of light within a wide frequency range centered at $\omega_{\rm c} = 2\pi/\lambda = 4.1357$ eV,  known as the stopband. With the two Bragg mirrors separated by $\lambda/2$, this Bragg resonator supports a standing cavity mode at frequency $\omega_{\rm c}$. As shown in the transmission spectrum of the cavity geometry (black line in Fig. \ref{fig:setup}b), this cavity mode is completely isolated from any other cavity lineshape by $\sim 1$ eV. The zoom-in cavity-mode lineshape (Fig. \ref{fig:setup}c) reveals a narrow linewidth of $\Delta \omega_{\rm c} \sim 0.01$ eV, corresponding to a  quality factor of $Q = \omega_{\rm c} / \Delta \omega_{\rm c} \sim 4\times 10^2$. This high-$Q$ cavity mode provides a realistic demonstration of the single-mode approximation commonly assumed in theoretical studies of polariton chemistry.

    We consider strong coupling when a layer of Lorentz medium with a width of $\lambda/2$ is confined at the center of the Bragg resonator. The  relative permittivity of the Lorentz medium takes the following form:
    \begin{equation}\label{eq:dielectric_lorentz}
        \varepsilon^{\ast}_{\rm r}(\omega) = \frac{\varepsilon(\omega)}{\varepsilon_{0}} = 1 + \frac{\sigma_{\rm L}\omega_{\rm L}^2}{\omega_{\rm L}^2 - \omega^2 - i\omega \gamma_{\rm L}} ,
    \end{equation}
    where $\varepsilon(\omega)$ denotes the material permittivity,   $\sigma_{\rm L}$, $\omega_{\rm L}$, and $\gamma_{\rm L}$ represent the oscillator strength, frequency, and dissipation rate of the Lorentz medium, respectively. The resonance condition, $\omega_{\rm L} =\omega_{\rm c}$, is  enforced throughout the manuscript, and  the dissipation rate is set as $\gamma_{\rm L} = 4.14\times 10^{-4}$ eV.  Simulation results for increased  values of $\gamma_{\rm L}$ are provided in the supplementary material.
    
    Fig. \ref{fig:setup}b presents the linear polariton spectra under different Rabi splitting $\Omega_{N}$ values. The Rabi splitting is tuned by varying the oscillator strength of the Lorentz medium as $\sigma_{\rm L} = 0.005m$, with $m=1,2,3,4$ represented by colors ranging from  dark red to orange, respectively. The corresponding Rabi splitting, indicated in the legend, increases proportionally to $\sqrt{\sigma_{\rm L}}$. This linear trend is shown more clearly in the inset of Fig. \ref{fig:setup}c. 

    \subsection{Spontaneous emission}

    When an electronic TLS is placed at the center of the Bragg resonator, Fig. \ref{fig:setup}d illustrates the relaxation dynamics of the TLS excited-state population in the absence of external pulse pumping. This TLS\red{, which is only weakly coupled to the cavity mode,} is initialized in a coherent state $\ket{\psi} = \sqrt{9/10}\ket{g} + \sqrt{1/10}\ket{e}$  with a frequency $\omega_{\rm TLS}=3.9404$ eV. The TLS population relaxation is driven by spontaneous emission, which can be  semiclassically described by the radiative self-interaction between the TLS and the self-emitted classical electric field. \cite{Jaynes1963,Li2018Spontaneous}
    When the TLS is placed in the Bragg resonator without the Lorentz medium (black), its population relaxation is strongly suppressed compared to the case in which it is placed in vacuum (cyan). This suppression arises from the Purcell effect: \cite{Purcell1995} since the TLS is off-resonant with the cavity mode supported by the Bragg resonator ($\omega_{\rm c} = 4.1357$ eV), the emitted electric field from the TLS experiences destructive interference, resulting in a reduced spontaneous emission rate.

    When the Lorentz medium is included to form strong coupling with the Bragg resonator, with parameters $\omega_{\rm L} = \omega_{\rm c}$ and  $\sigma_{\rm L} = 0.02$, the resulting lower polariton (LP) matches exactly with the TLS frequency, $\omega_{\rm LP} = \omega_{\rm TLS} = 3.9404$ eV. Consequently, as shown in Fig. \ref{fig:setup}d, the TLS decay dynamics are significantly enhanced (orange) compared to the vacuum case. This polariton-enhanced radiative decay can be attributed to the Purcell enhancement effect: the formation of strong coupling causes the cavity mode to oscillate at both the LP and upper polariton (UP) frequencies. When the TLS frequency aligns with either the LP or UP frequency, the emitted electric field experiences constructive enhancement, leading to an increased spontaneous emission rate. Hereafter, we refer to this Purcell effect under strong coupling as the \textit{polariton-induced Purcell effect}.
    
    The predicted spontaneous emission lifetime for the TLS in vacuum is on the order of tens of ps. This lifetime is significantly shorter than the typical spontaneous emission lifetime of organic molecules, which is on the order of a few ns. \cite{Berezin2010} This difference arises from the use of a large transition dipole moment for the TLS,  $\mu_{12} = 56.9$ Debye, which exceeds realistic values by  more than an order of magnitude. Such a large $\mu_{12}$ is employed to reduce the simulation time steps. Although the applied $\mu_{12}$ is artificially large, the presence of the TLS does not alter the linear polariton spectrum; see Fig. \ref{fig:LinearSpectroscopy} in the Appendix. This invariance indicates that the TLS itself is only weakly coupled to the cavity, consistent with the experimental limit. As the presence of the TLS does not alter the photonic density of states inside the cavity, according to the Fermi's golden rule,  spontaneous emission rates are always proportional to $|\mu_{12}|^2$ regardless of the cavity structures. Consequently, using a large $\mu_{12}$  increases the spontaneous emission rates both outside and inside the cavity proportionally. In other words, the relative cavity effects on spontaneous emission, such as the Purcell factor (defined as the ratio of the spontaneous emission rates  inside versus outside cavity), can still be accurately captured with our approach. Note that if $\mu_{12}$ were large enough to meaningfully alter the linear polariton spectrum --- thereby changing the photonic density of states due to the presence of the TLS --- the simulated cavity effects on spontaneous emission would not be reliable.

    \begin{figure}[h]
	    \centering
	    \includegraphics[width=0.95\linewidth]{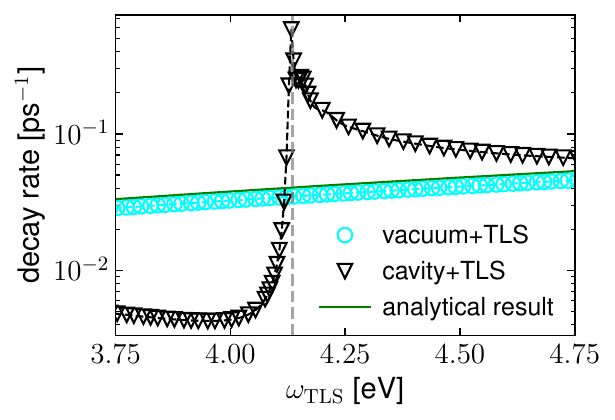}
	    \caption{
        Fitted TLS decay rate as a function of its frequency $\omega_{\rm{TLS}}$ under two conditions: inside an empty Bragg resonator (black triangles) versus in vacuum (cyan circles). Inside the cavity, the maximal spontaneous emission rate occurs when $\omega_{\rm{TLS}}$ is at resonance with the cavity frequency  $\omega_{\mathrm{c}}=4.1357$ eV (vertical dashed line). The asymmetric decay rate in the cavity may arise from the Fano interference effect.}
	    \label{fig:Purcell_effect}
    \end{figure}

    \subsubsection{Purcell effects under weak coupling}
    
    To further compare the polariton-enhanced Purcell effect with the conventional Purcell effect under weak coupling, we first revisit the conventional Purcell effect through simulations. When the TLS is placed in vacuum, Fig. \ref{fig:Purcell_effect} shows the fitted TLS decay rate (cyan circles) as a function of $\omega_{\rm{TLS}}$. The numerical result agrees with the analytical rate $k_{\rm sc} = \rho_{11}(0)|\mu_{12}|^2\omega_{\rm TLS}^2/2\hbar \epsilon_0 c^2$ (solid green line), as derived in Eq. \ref{eq:SE_sc_rate} of the Appendix.  When the TLS is confined within the empty Bragg resonator (black triangles), a strong rate enhancement is observed when $\omega_{\rm TLS}$ approaches resonance with the cavity mode at $\omega_{\rm c} = 4.1357$ eV.

    \begin{figure*}
	    \centering
	    \includegraphics[width=1.0\linewidth]{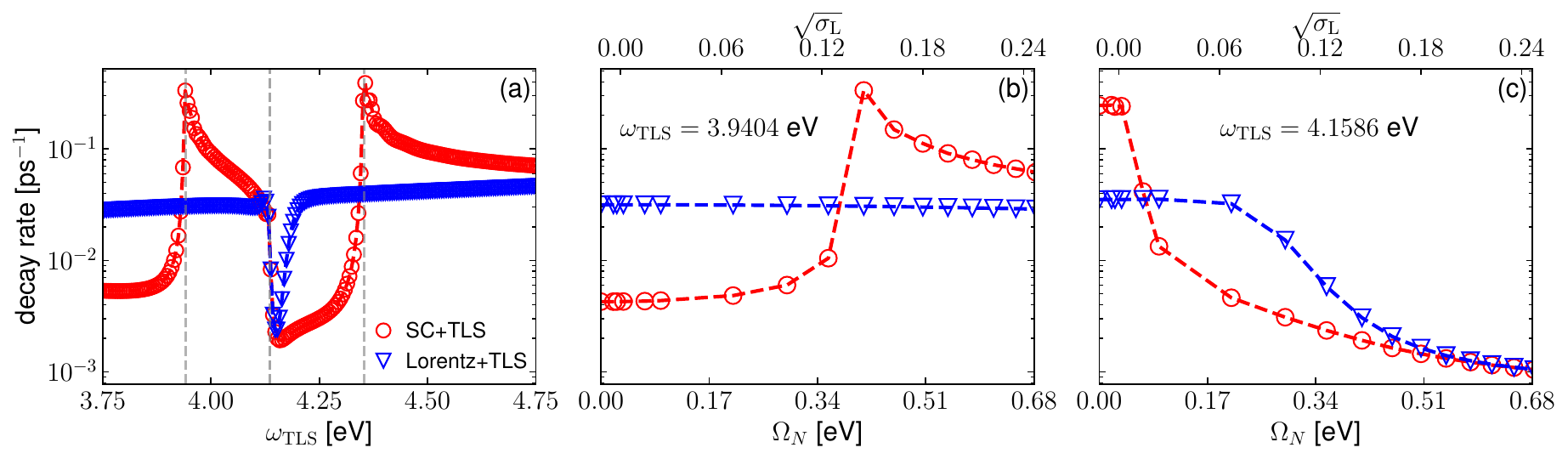}
	    \caption{Parameter dependence of the polariton-induced Purcell effect. (a) TLS decay rate as a function of $\omega_{\rm{TLS}}$ when the TLS is placed within the Lorentz medium under two conditions: inside (red dots) versus outside the cavity (blue triangles). The three vertical dashed gray lines (from left to right) indicate the LP frequency $\omega_{\rm{LP}} = 3.9404$ eV, $\omega_{\rm{c}}=\omega_{\rm{L}} =  4.1357$ eV, and the UP frequency $\omega_{\rm{UP}} = 4.3560$ eV, respectively. (b) TLS decay rate as a function of $\sqrt{\sigma_{\rm L}}$ inside versus outside the cavity. The corresponding Rabi splitting values under strong coupling are also shown on the $x$-axis.  The TLS frequency is set as $\omega_{\rm{TLS}} = 3.9404$ eV, corresponding to the LP frequency  in part (a). (c) Similar graph as part (b), except that the TLS frequency is set to  $\omega_{\rm{TLS}} = 4.1586$ eV, near resonance with $\omega_{\rm c} = \omega_{\rm{TLS}} = 4.1357$ eV. Forming polaritons suppresses (or enhances) the TLS radiative decay rate when $\omega_{\rm{TLS}}$ is near resonance with the cavity (or polariton) frequency.}
	    \label{fig:radiative_decay_dependence}
    \end{figure*} 
    
    An intriguing result in Fig. \ref{fig:Purcell_effect} is the asymmetric behavior of the decay rate under \red{weak} coupling as the TLS frequency deviates from $\omega_{\rm c}$. For $\omega_{\rm TLS} > \omega_{\rm c}$, increasing the positive detuning ($\omega_{\rm TLS} - \omega_{\rm c}$) yields a modestly suppressed spontaneous emission rate, which remains larger than the corresponding vacuum result. In contrast, for $\omega_{\rm TLS} < \omega_{\rm c}$, increasing the negative detuning ($\omega_{\rm c} - \omega_{\rm TLS}$)  leads to a strongly suppressed spontaneous emission rate, reaching one-order-of-magnitude below the vacuum limit. This asymmetry can be attributed to the Fano interference effect, which arises when two scattering pathways --- the discrete cavity resonance scattering (along the $x$-axis)  and the background scattering into a continuum (along the $y$-axis) --- coexist and interfere with each other. \cite{Fano1961,Yamaguchi2021} \red{This asymmetry cannot be captured when the EM environment is not explicitly simulated and the cavity is modeled as a single harmonic oscillator; see Appendix \ref{appendix:single-mode} for details.}

    \subsubsection{Polariton-induced Purcell effects}

    Building on the understanding of the Purcell effect under weak coupling, we now focus on the polariton-induced Purcell effect. Fig. \ref{fig:radiative_decay_dependence}a shows the fitted TLS decay rate as a function of the TLS frequency under strong coupling (red dots) compared to the case outside the cavity (blue triangles). In both scenarios, the TLS is placed at the center of  the Lorentz medium with $\sigma_{\rm L} = 0.02$. 
    
    Outside the cavity, the decay rate exhibits a sharp negative peak when the TLS frequency approaches resonance with the Lorentz medium frequency (blue triangles). This suppression can be explained by the dielectric properties of the Lorentz medium: near resonance, the dielectric constant $\varepsilon(\omega\approx\omega_{\rm L})$ becomes extremely large; consequently, the amplitude of the  electric field emitted by the TLS is significantly reduced (as $\vE = \vD/\varepsilon$, where $\vE$ and $\vD$ denote the electric field and the displacement field, respectively). With a diminished emitted electric field amplitude, the radiative self-interaction of the TLS, which drives its population decay, is strongly suppressed.

    Under strong coupling, the TLS decay rate in Fig. \ref{fig:radiative_decay_dependence}a exhibits two sharp resonance enhancement peaks when the TLS frequency matches either the LP or UP frequency (indicated by the vertical dashed lines at two sides). As the TLS frequency deviates from each polariton frequency, the radiative decay rate decreases asymmetrically. This asymmetric behavior resembles that observed under weak coupling (Fig. \ref{fig:Purcell_effect}), as if the cavity mode were split to two independent cavity modes at frequencies $\omega_{\rm LP}$ and $\omega_{\rm UP}$. Notably, when the TLS frequency is tuned near resonance with $\omega_{\rm L} = \omega_{\rm c}$ (the middle vertical dashed line), the decay rate is strongly suppressed compared to the outside-cavity result, particularly when $\omega_{\rm TLS}$ exceeds  $\omega_{\rm L} = \omega_{\rm c}$. The maximal decay rate suppression occurs at $\omega_{\rm TLS} = 4.1586$ eV inside the cavity, a value 0.6\%  higher than $\omega_{\rm c} = \omega_{\rm L} = 4.1357$ eV. Such a small red shift may potentially arise from the semiclassical inclusion of the self-polarization term in the Hamiltonian, which leads to a Lamb-shift-like blueshift when the TLS  experiences the contact interactions with the surrounding Lorentz medium. For a more detailed discussion of the the self-polarization term, or the longitudinal light-matter interaction term,  see Sec. \ref{sec:methods}.  

    Fig.  \ref{fig:radiative_decay_dependence}b further illustrates the TLS decay rate versus the Rabi splitting when $\omega_{\rm TLS} =  3.9404$ eV. In practice, the Rabi splitting is varied by tuning the oscillator strength of the Lorentz medium $\sigma_{\rm L}$. Under weak coupling (when the Rabi splitting is very small), since $\omega_{\rm TLS}$ is not resonant with the cavity frequency, the TLS decay rate inside the cavity (red circles) is smaller than that outside the cavity (blue triangles) due to the Purcell suppression effect. As the Rabi splitting increases, the radiative decay rate is gradually enhanced. The maximal enhancement occurs when $\Omega_{N} =  0.4127$ eV ($\sigma_{\rm L} = 0.02$), which corresponds to the condition when $\omega_{\rm TLS}$ matches the LP frequency. Beyond this point, as the Rabi splitting continues to grow and the TLS deviates from the LP frequency, the radiative decay rate becomes reduced.
    
    \begin{figure*}
	    \centering
	    \includegraphics[width=1\linewidth]{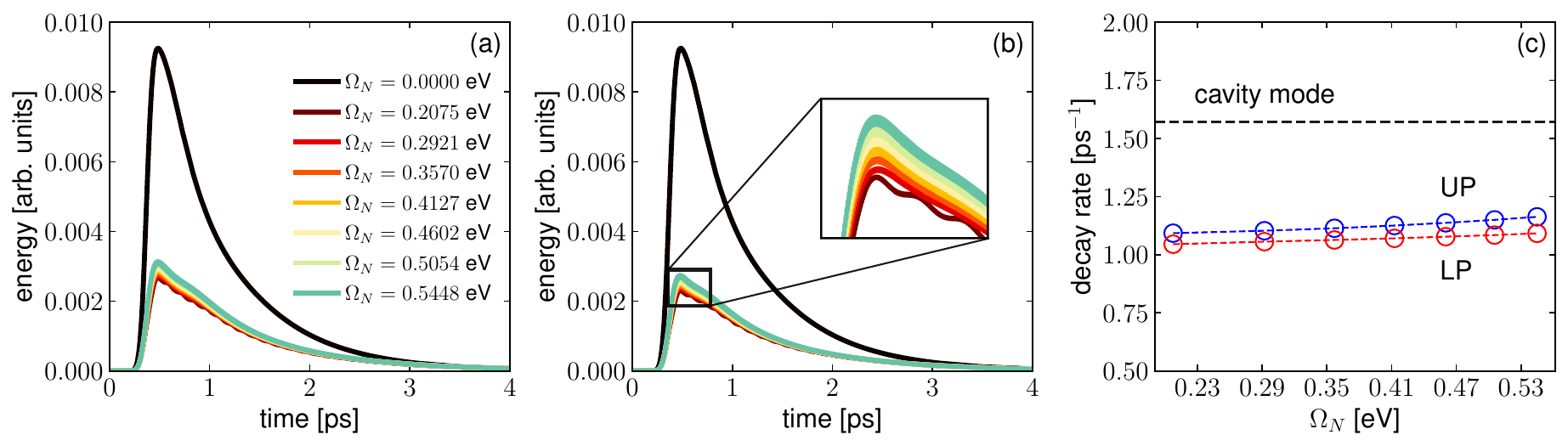}
	    \caption{Simulated polariton energy relaxation dynamics under  external Gaussian pulse excitation of (a) the LP and (b) the UP. Colored lines (from brown to green) represent the strong coupling system under different Rabi splittings, as labeled in the legend. The empty cavity case, where the Gaussian pulse resonantly excites the cavity frequency, is also included (black line). The polariton (or cavity) energy is calculated by integrating the EM field energy density confined between the two Bragg mirrors. A moving average with a duration of 20 fs is applied on the time-dependent signals. (c) Fitted polariton decay rates versus the Rabi splitting corresponding to parts (a) and (b). The UP (blue circles) and LP (red circles) decay rates are insensitive to the Rabi splitting. For comparison, the cavity decay rate in the absence of the Lorentz medium is also shown as the horizontal dashed black line.        }
	    \label{fig:polariton_relaxation}
    \end{figure*}
    
    For the TLS  at frequency $\omega_{\rm TLS} = 4.1586$ eV $\approx \omega_{\rm L} = \omega_{\rm c}$, Fig. \ref{fig:radiative_decay_dependence}c shows that the outside-cavity decay rate decreases when the oscillator strength $\sigma_{\rm L}$ increases. This reduction, as discussed around Fig. \ref{fig:radiative_decay_dependence}a, can be attributed to the increased dielectric constant of the Lorentz medium at $\varepsilon(\omega=\omega_{\rm TLS})$ as the oscillator strength $\sigma_{\rm L}$ amplifies. Inside the cavity, the TLS decay rate  under weak coupling is larger than the outside-cavity result  due to the Purcell enhancement effect. However, as the Rabi splitting increases, the cavity gradually suppresses the TLS decay rate below the outside-cavity limit. At the limit of a very large Rabi splitting, the inside- versus outside-cavity decay rates converge. This convergence  can be explained as follows: when the oscillator strength of the Lorentz medium becomes very large, the high dielectric constant at $\varepsilon(\omega=\omega_{\rm TLS})$ strongly suppresses the emitted electric field from the TLS before it reaches the cavity mirrors; as a result, the cavity effect on the spontaneous emission becomes negligible.

    \subsection{Polariton relaxation dynamics}

    \red{In the results presented above, the polariton-induced Purcell effect was observed when the excited TLS acted as the sole emitter of the EM field. In other words, no external pulse was applied on the system. Here,} Fig. \ref{fig:polariton_relaxation} examines the time-resolved dynamics of the strong coupling system under an external Gaussian pulse excitation. In this simulation, the TLS dynamics are excluded, and our approach reduces to the conventional FDTD method.
    
    As shown in Fig. \ref{fig:polariton_relaxation}a, when the Gaussian pulse resonantly excites the cavity in the absence of the Lorentz medium (black line), the integrated EM field energy inside the cavity  initially increases under the pulse excitation and  then gradually decreases due to the cavity loss. If the Lorentz medium is included to form strong coupling, as shown in Fig. \ref{fig:polariton_relaxation}a (or Fig. \ref{fig:polariton_relaxation}b), resonant excitation of the LP (or the UP) reduces the EM field energy accumulation inside the cavity. This occurs because the cavity structure, rather than the Lorentz medium,  exhibits a strong optical response to external field excitation; with the formation of polaritons, because each polariton mode comprises roughly half of the cavity component, its response to the external field is greatly reduced compared to the case of an empty cavity. Assuming that the effective transition dipole moment of each polariton, $\mu_{\rm P}$, is half that of the cavity, $\mu_{\rm c}$, and given that energy absorption is proportional to $|\mu|^2$, the maximal transient energy absorption of each polariton should be approximately one-quarter of that of an empty cavity. This estimation is consistent with Fig. \ref{fig:polariton_relaxation}.
    
    Interestingly, under strong coupling, the EM field energy accumulation can be slightly amplified by exciting the polaritons with a larger Rabi splitting. This trend is consistent with the Rabi splitting dependence of the linear polariton peak strength shown in Fig. \ref{fig:setup}c. This behavior is likely attributed to the realistic cavity structure included in the simulation, i.e., the polariton peaks closer to the edge of the Bragg stopband may exhibit more absorptive characteristics. Note that in Figs. \ref{fig:polariton_relaxation}a and \ref{fig:polariton_relaxation}b, the polariton relaxation dynamics for $0< \Omega_{\text{N}} < 0.2075$ eV are not shown. Within this range of small Rabi splittings, especially when the two polariton peaks are not well separated, using the Gaussian pulse may simultaneously excite both the UP and LP.

    Fig. \ref{fig:polariton_relaxation}c further provides the fitted polariton decay rates obtained from the time-dependent trajectories. The polariton relaxation rates remain mostly unchanged across different Rabi splittings. This invariance contrasts with more realistic calculations of polariton relaxation dynamics, where polariton relaxation rates can be strongly suppressed as the Rabi splitting increases.\cite{Groenhof2019,Li2021Relaxation} In these advanced simulations, the suppression occurs because the polariton dephasing into dark modes becomes less likely when the Rabi splitting increases. 
    Importantly, the absence of this Rabi-splitting dependence in polariton decay rates is robust against the variations in $\gamma_{\rm L}$, the Lorentz medium dissipation rate; see the supplementary material for simulation results with increased $\gamma_{\rm L}$ values.  The lack of Rabi-splitting dependence in polariton decay rates stems from modeling the confined molecules as simple dielectric functions (such as the Lorentz medium), an approximation in both the conventional FDTD approach and our reduced semiclassical electrodynamics approach. This limitation occurs because simple dielectric functions cannot capture inter- or intramolecular interactions, whereas polariton dephasing into dark modes is driven by these molecular interactions.\cite{Groenhof2019,Li2021Relaxation} 

\begin{figure*}
	    \centering
	    \includegraphics[width=0.8\linewidth]{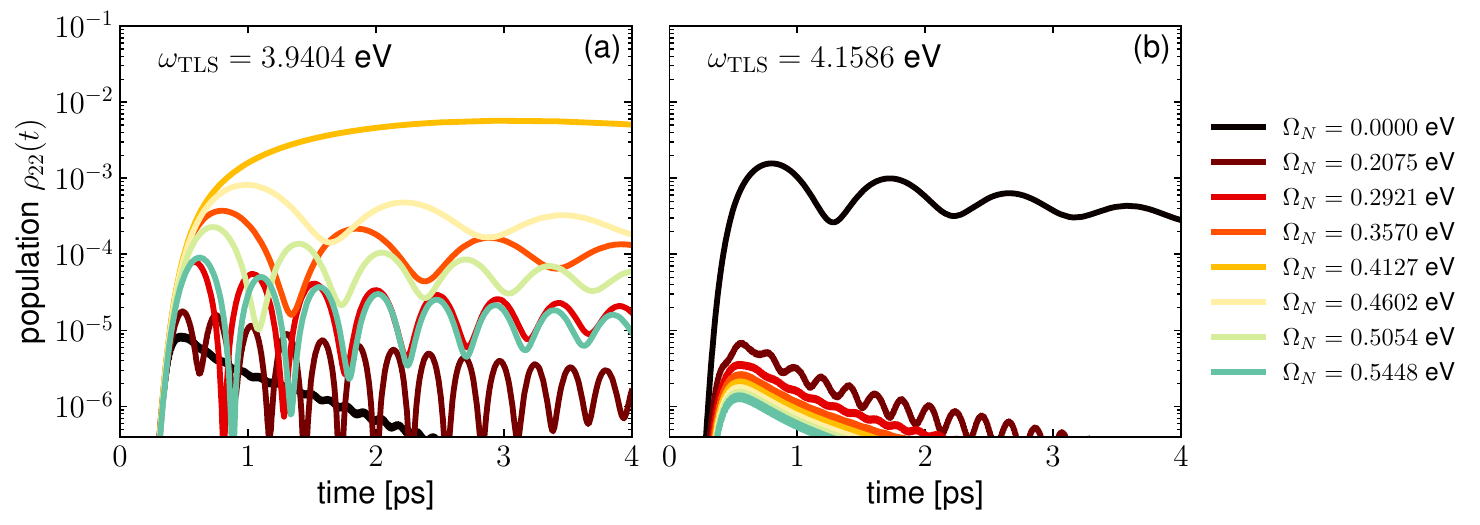}
	    \caption{TLS excited-state population dynamics after  the LP excitation corresponding to Fig. \ref{fig:polariton_relaxation}a. Two conditions are compared:  (a) the TLS frequency set to $\omega_{\mathrm{TLS}}=$3.9404 eV, matching the LP frequency when the Rabi splitting is $\Omega_{N} = 0.4127$ eV (orange line);  (b) the TLS frequency set to $\omega_{\mathrm{TLS}}=$4.1586 eV, near resonance with the bare cavity frequency (black line). Under strong coupling, the TLS excited-state population response is strongly enhanced (or suppressed) when its frequency matches the LP frequency (or the bare cavity frequency).}
	    \label{fig:TLS_Pulse}
    \end{figure*}

        \subsection{Driven TLS dynamics under strong coupling}

    When the TLS dynamics are include in the simulations,  Fig. \ref{fig:TLS_Pulse} shows the excited-state population dynamics of the TLS driven by the LP excitation corresponding to Fig. \ref{fig:polariton_relaxation}a. When the TLS frequency is set as $\omega_{\rm TLS}$ = 3.9404 eV (Fig. \ref{fig:TLS_Pulse}a), matching the LP frequency for a Rabi splitting of $\Omega_{\rm N} = 0.4127$ eV (orange line), resonant excitation of this LP yields the maximal TLS excited-state response. As the Rabi splitting deviates from this value, the TLS excited-state response is significantly suppressed.   
    
    When the TLS frequency is set to $\omega_{\rm TLS}$ = 4.1586 eV (Fig. \ref{fig:TLS_Pulse}b), near resonance with $\omega_{\rm c} = \omega_{\rm L}$ (black line), the maximal TLS excited-state response occurs only when the empty cavity is resonantly excited. When the Lorentz medium is included to form strong coupling (from brown to cyan lines), resonant excitation of the LP  yields a strongly suppressed TLS excited-state response as the Rabi splitting increases. The polariton-enhanced or suppressed TLS driven dynamics in Fig. \ref{fig:TLS_Pulse} align with the polariton-induced Purcell effect of spontaneous emission   in Fig. \ref{fig:radiative_decay_dependence}.

    Throughout this manuscript, the dissipation rate of the Lorentz medium has been set as $\gamma_{\rm L} = 4.14\times 10^{-4}$ eV, corresponding to a dissipation lifetime of $1/2\pi \gamma_{\rm L} \sim 1.6$ ps. Since this lifetime is longer than the photonic lifetime captured in Fig. \ref{fig:polariton_relaxation} ($\sim 0.6$ ps), in the supplementary material we also provide two sets of simulation data analogously to Figs. \ref{fig:setup}--\ref{fig:TLS_Pulse} for cases when $\gamma_{\rm L}$ is increased by a factor of 10 and $10^2$, respectively. Under these large dissipation rates, the polariton effects on TLS dynamics remain qualitatively consistent with those reported in Figs. \ref{fig:setup}--\ref{fig:TLS_Pulse}. The only notable discrepancy occurs when $\gamma_{\rm L}$ significantly exceeds the cavity decay rate ($\gamma_{\rm L} = 4.14\times 10^{-2}$ eV, corresponding to a lifetime of $\sim 0.02$ ps). In this strongly dissipative regime, the TLS radiative decay rate under strong coupling is always larger that that outside the cavity, even when the TLS  is near resonance with the bulk molecules forming strong coupling.

    \red{Lastly, we briefly discuss how inaccuracies in describing the dependence of the polariton dephasing lifetime on the Rabi splitting (Fig. \ref{fig:polariton_relaxation}c) may affect other results presented in this manuscript. Due to this limitation, the simulation results for the driven TLS dynamics under external polariton pumping (Fig. \ref{fig:TLS_Pulse}) may not be entirely accurate, as the excited polariton is expected to transfer energy to the TLS via the polariton dephasing mechanism. However, the polariton-induced Purcell effect results shown in Fig. \ref{fig:radiative_decay_dependence} are likely less affected by this limitation, because the radiative relaxation of the excited TLS is governed by the photonic density of states, which our current simulation accurately captures. }

    \section{Discussion: Connecting to Experiments}\label{sec:discussion}

    Our calculations provide a detailed study of the polariton-induced Purcell effect by analyzing a quantum impurity at various frequencies. This effect arises from the altered photonic density of states under strong coupling, thus surviving in the large $N$ limit. While the  enhancement of radiative decay rates by polaritons has been studied experimentally, \cite{Zhao2024} our calculations suggest that polaritons can also suppress radiative decay rates when the quantum impurity is near resonance with the bulk molecules forming strong coupling, at which frequency the photonic density of states is greatly reduced compared to that under weak coupling. This is in contrast to the early suggestion by Hutchison \textit{et al} \cite{Hutchison2012} that cavity leakage of the LP may enhance the radiative decay rates of an ensemble of identical molecules. One direct experimental consequence of the polariton-suppressed Purcell effect is, for the case of $N$ identical molecules forming electronic strong coupling, after external pumping, the strong coupling system may exhibit prolonged photoluminescence signals  on a timescale much longer than the polariton lifetime.
    
    Our simulation results may facilitate understanding  certain experimental observations in polariton chemistry, particularly those exploring whether local molecular processes can be altered by strong coupling. Our simulations highlight two mechanisms that may explain  polariton-induced photochemistry: the reduced EM absorption under strong coupling, and the polariton-induced Purcell effect. The former mechanism has been discussed in the literature, \cite{Thomas2024} but this mechanism can only explain the polariton-induced rate suppression \cite{Hutchison2012} rather than the acceleration \cite{Zeng2023}. The polariton-induced Purcell effect, in contrast, may result in both rate suppression and acceleration, with the assumption that an enhanced radiative process competes nonradiative process, such as chemical reactions. Specifically, when the impurities (e.g., molecules undergoing chemical reactions) oscillate near the polariton frequencies, because the radiative process is enhanced due to the polariton-induced Purcell effect, the corresponding nonradiative decay rate can be suppressed. Conversely, when the impurity frequency is near resonance with the bulk molecular frequency, because  the polariton-induced Purcell effect often predicts suppressed radiative processes, nonradiative processes such as reactions may be accelerated under strong coupling.
    
    Beyond this qualitative analysis, it is important to consider the timescale on which the polariton-induced Purcell effect may influence photochemical processes. Given that this effect may enhance or suppress  radiative decay rates by an order of magnitude, and the radiative decay rates for organic molecules in electronically excited states are typically on the order of a few ns, \cite{Berezin2010} the polariton-induced Purcell effect has the potential  to alter photochemical reaction processes whose rate-determining nonradiative pathways are on the timescale ranging from hundreds of ps to tens of ns. 

    This analysis does not preclude other possible mechanisms for polariton-altered photochemistry. For instance, during polariton relaxation, the excited polaritons may transfer energy to other molecular degrees of freedom and create a unique molecular excited-state distribution in the dark states. This process could ultimately   alter photochemistry on a timescale much longer than the polariton lifetime, as discussed in Refs. \citenum{Chen2022,Zeng2023} for possible mechanisms. However, because our approach models most molecules as dielectric functions and does not include the dark-mode degrees of freedom, it is not suited to investigate this nontrivial polariton effect. Future studies, potentially employing alternative simulation approaches, should explore this possibility. 

    \section{Conclusion}

    To summarize, we have simulated the local TLS dynamics under collective strong coupling formed by a Lorentz medium confined in a realistic 2D Bragg resonator. By self-consistently solving the coupled Maxwell--Schr\"odinger equations, we have made three major observations.
    \begin{itemize}
        \item[(i)] Due to the polariton-induced Purcell effect, the spontaneous emission of the TLS can be enhanced under collective strong coupling when the TLS frequency is near resonance with the polariton frequency. This effect is very similar to the conventional Purcell effect under weak coupling, as if strong coupling splits the original  cavity mode to two independent cavity modes oscillating at the LP and UP frequencies. By contrast, when the TLS frequency is near resonance with the bulk molecular frequency, the spontaneous emission  of the TLS can be strongly suppressed under strong coupling, provided that the Lorentz medium is not excessively dissipative. 
        \item[(ii)] When the strong coupling system is driven by an external Gaussian pulse,  our reduced semiclassical electrodynamics approach captures suppressed, Rabi-splitting-dependent EM energy absorption compared to  weak coupling. The corresponding local TLS response is also governed by the polariton-induced Purcell effect. 
        \item[(iii)] While the above two observations suggest that polaritons behave similarly to pure cavity modes with modified frequencies and lineshapes, our simulations reveal a significant limitation of modeling most molecules as simple dielectric functions: the inability to capture the Rabi-splitting dependence of polariton dephasing rates into the dark modes. 

    \end{itemize}

    As discussed in Sec. \ref{sec:discussion}, the polariton-induced Purcell effect may offer a unique perspective to understand the origin of polariton-induced photochemistry in the large $N$ limit. From a theoretical perspective, our simulations emphasize the significance of incorporating a realistic cavity structure, or a realistic photonic bath. This is crucial because the spontaneous emission rate of molecules can only be accurately described with a photonic bath under the wide-band approximation \cite{Nitzan2006}, rather than by considering only a single or a few cavity modes. Given  the computational efficiency of our reduced semiclassical electrodynamics approach, future work should aim to address its limitations, such as the inaccurate description of polariton dephasing rates into dark modes. With these improvements, this approach may potentially provide a robust framework for simulating local molecular processes under collective strong coupling. 

    \section{Supplementary Material}
    Two sets of simulation data analogously to Figs. \ref{fig:setup}--\ref{fig:TLS_Pulse} for cases when the dissipation rate of the Lorentz medium $\gamma_{\rm L}$ is set to $4.14\times 10^{-3}$ eV and $4.14\times 10^{-2}$ eV, respectively.
    
\section{Acknowledgments}
    This material is based upon the work supported by the start-up funds from the University of Delaware Department of Physics and Astronomy. This research is supported in part through the use of Information Technologies (IT) resources at the University of Delaware, specifically the high-performance computing resources.  

    \section{Data Availability Statement}
    The code and data that support the findings of this study are available at the following Github repository: \url{https://github.com/TaoELi/semiclassical_electrodynamics}.

\appendix

    \setcounter{table}{0}
	\setcounter{equation}{0}
	\renewcommand{\theequation}{A\arabic{equation}}

	\maketitle
	
    
\section{Spontaneous emission rates from the Fermi's golden rule}\label{sec:FGR}

According to the Fermi's golden rule, the spontaneous emission rate for an electronically excited TLS in vacuum can be calculated as: 
\begin{subequations}
    \begin{equation}
    k_{\rm QM}=\frac{2\pi}{\hbar^{2}}\sum_{\vec{k}^{\prime},\vec{s}} \frac{\hbar \omega_{k^{\prime}}}{2\epsilon_0 V}\left| \vec{\mu}_{12} \cdot \vec{\epsilon}_{\vec{k}^{\prime}, \vec{s}} \right|^2 \delta(\omega_{\rm TLS} - ck^{\prime}) . 
\end{equation}
Here, $\vec{k}'$, $\vec{s}$, and $\omega_{k'}$ denote the wave vector, polarization direction, and frequency of the photons, respectively; $\vec{\epsilon}_{\vec{k}^{\prime}}$ represents the unit vector along the photon polarization direction; $V$ denotes the quantization volume of the photons; and $\vec{\mu}_{12}$ is the transition dipole vector of the TLS.

In our reduced semiclassical electrodynamics simulations,   the transition dipole moment of the TLS is  assumed to orient along the $z$-direction, perpendicular to the $x$-$y$ simulation plane. Additionally, only the electric field polarized along the $z$-direction (the transverse electric mode, or the TE mode) is considered, so it is unnecessary to consider two possible polarization directions. With these considerations in mind, $\vec{\epsilon}_{\vec{k}^{\prime}} = \mathbf{e}_z$ and  $\vec{\mu}_{12} = \mu_{12} \mathbf{e}_z$, allowing us to express the TLS decay rate as:
\begin{equation}
    \begin{aligned}
    k_{\rm QM} &= \frac{2\pi \left| \mu_{12} \right |^2}{\hbar^{2}}\sum_{\vec{k}^{\prime}} \frac{\hbar \omega_{k^{\prime}}}{2\epsilon_0 V} \delta(\omega_{\rm TLS} - ck^{\prime})  \\
    &= \frac{2\pi \left| \mu_{12} \right |^2}{\hbar^{2}}
    \int_0^{2\pi} d\phi \int_{0}^{+\infty} k' d k'  \frac{L_x L_y}{(2\pi)^2} \frac{\hbar \omega_{k^{\prime}}}{2\epsilon_0 V}  \\
     & \ \  \ \times \delta(\omega_{\rm TLS} - ck^{\prime}) \\ 
    &= \frac{|\mu_{12}|^2\omega_{\rm TLS}^2}{2\hbar\epsilon_0 c^2} .
\end{aligned}
\end{equation}
\end{subequations}
Here, $V = L_x L_y$, where $L_x$ and $L_y$ denote the lengths along the $x$- and $y$-directions, respectively. 

In the semiclassical electrodynamics approach,  the electric field is treated classically, so vacuum fluctuations of photons cannot be properly described. Consequently, the semiclassical TLS spontaneous emission rate is proportional to $\rho_{11}(0)$, the initial ground-state population of the TLS: \cite{Jaynes1963,Li2018Spontaneous}
\begin{equation}\label{eq:SE_sc_rate}
    k_{\rm sc} = \rho_{11}(0) k_{\rm QM} = \rho_{11}(0) \frac{|\mu_{12}|^2\omega_{\rm TLS}^2}{2\hbar\epsilon_0 c^2} .
\end{equation}
Hence, to accurately describe the spontaneous emission rate within the reduced semiclassical electrodynamics approach, we set $\rho_{11}(0) = 0.9$ when performing calculations.

\setcounter{equation}{0}
	\renewcommand{\theequation}{B\arabic{equation}}

\section{Units conversion in simulations}\label{sec:units}

In conventional FDTD simulations, $\epsilon_0 = c_0 = 1$ is used. For propagating the TLS dynamics, we also set $\hbar = 1$. By also defining the units of time as $[T] = 0.1$ fs = 1 MEEP units (mu) in our simulations, we can use these four quantities to determine the SI values of all other parameters. For instance, the SI value of angular frequency $\omega$ can be expressed as:
\begin{equation}
    [\omega] = \frac{2\pi}{[T]} = 41.357 \mathrm{\ eV} = 1 \mathrm{\ mu} .
\end{equation}
It is worth noting that the MEEP code uses rotational frequency $f = \omega/2\pi$ instead of $\omega$ when defining the frequency of the Lorentz medium or the pulse. For example, if  the TLS (angular) frequency is set as $\omega_{\rm TLS} = 0.1$ mu, and we wish to use a pulse (defined in MEEP) to resonantly excite the TLS, the (rotational) frequency of this pulse must be set to $f = 0.1 / 2\pi$ mu in the MEEP code.

The  SI value of the length is taken as:
    \begin{equation}
        [X] = \frac{1}{[f]} = \frac{c_0 [T]}{2\pi}  = 4.771 \text{\ nm} = 1 \text{\ mu} .
    \end{equation}
Importantly, the SI  value of the dipole moment $\mu_{12}$ is
    \begin{equation}
    \begin{aligned}
        [\mu_{12}] &= [\text{charge}]\cdot [X] = \sqrt{c_0\epsilon_0\hbar}\cdot [X] \\
        &= 6.328\times 10^{-27}\text{ C$\cdot$ m} = 1.8970\times 10^{3} \text{ Debye} = 1 \text{ mu} .
    \end{aligned}
    \end{equation}

\setcounter{equation}{0}
	\renewcommand{\theequation}{C\arabic{equation}}
 \section{Simulation details}\label{sec:simu_details}

    \begin{figure}
	    \centering
	    \includegraphics[width=0.9\linewidth]{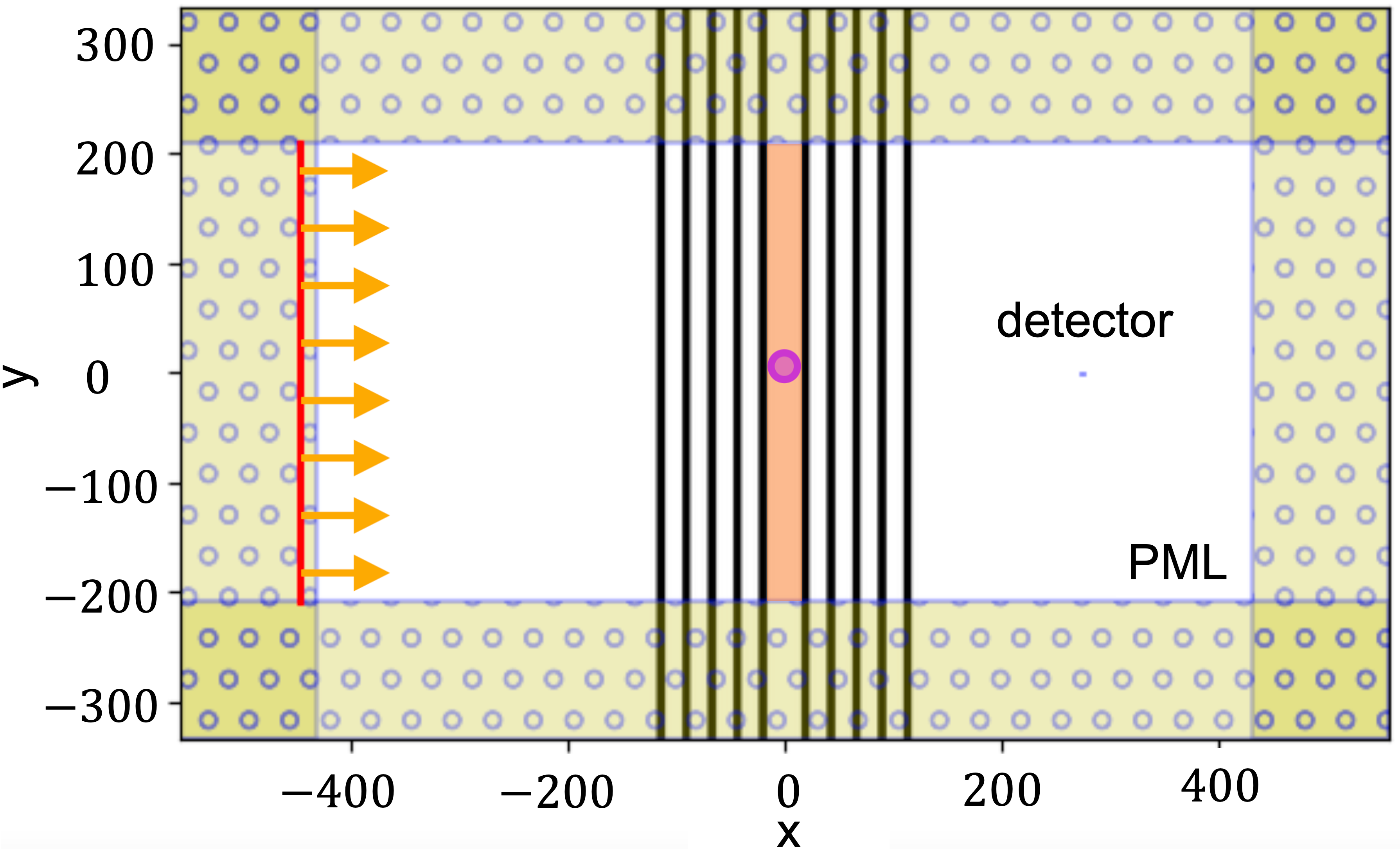}
	    \caption{Detailed geometry of the 2D simulation cell. The electronic TLS (purple circle) is placed at the origin of the cell,  corresponding to the center of the Lorentz medium (orange) located between the Bragg mirrors (black).  The source of the external pulse (red line) is placed at the boundary of the PML, serving as the origin of the plane wave (amber arrows). A spectra detector (blue) is located at the far right side of the Bragg resonator to compute the transmittance spectra. Spatial ticks are given in MEEP units (mu).
        }\label{fig:SI_SetUp}
    \end{figure}
    
    As mentioned in the main paper, the EM field was  propagated classically using the FDTD algorithm implemented in the MEEP package. The most relevant quantities were presented in SI units. For clarity, however, the simulation details provided here included both SI units and mu. 
    
    As illustrated in Fig. \ref{fig:SI_SetUp}, the system was simulated using a cell size of $5.32 \times 3.19\; \mu \mathrm{m}^{2}$ $(1115.27 \times 669.16\; \mathrm{mu}^{2})$. This simulation cell was sufficiently large to ensure numerical convergence. The EM field near the boundary of the simulation cell was absorbed using the perfectly matched layers (PML) technique. The PML width was set as 0.6 $\mu$m ($40\pi$ mu). The spatial and temporal  resolutions of the simulation were set to  $\Delta x = 7.49$ nm ($\pi/2$ mu) and $\Delta t = 0.0785$ fs ($\pi/4$ mu), respectively. \red{Additionally, the real-time propagation of the quantum TLS was performed at the same time step as the Maxwell's equations.}
    
    As mentioned in the main paper, a 2D Bragg resonator was placed at the center of the simulation cell. The Bragg resonator was composed of a pair of parallel mirrors separated by a distance of $\lambda/2 = 149.89$ nm ($10\pi$ mu). Each mirror consisted of five layers of dielectric media, each with a refractive index $n = 2$.  The width of each dielectric layer was set to  $\lambda/8 = 37.47$ nm ($2.5\pi$ mu). The neighboring dielectric layers were separated by a distance of $\lambda/4 = 74.95$ nm ($5\pi$ mu) filled with air (with refractive index $n = 1$). 
    
    For forming strong coupling, a layer of Lorentz medium with a length of $\lambda/2$ was placed at the center of the Bragg resonator. The relative permittivity of the Lorentz medium was given by Eq. \eqref{eq:dielectric_lorentz}. The parameters of the Lorentz medium were set as $\omega_{\mathrm{L}}=4.1357$ eV (0.1 mu) and $\gamma_{\mathrm{L}}=4.14\times 10^{-4}$ eV ($10^{-5}$ mu). The oscillator strength of the Lorentz medium was adjusted to vary the Rabi splitting, with a value of $\sigma_{\rm L} = 0.02$ most frequently used throughout the manuscript.

    Additionally, the electronic TLS was positioned at the origin, the center of the Lorentz medium. The polarization density of the TLS was defined in Eq. \eqref{eq:polarization_2d_method}. Numerically, the polarization density was expressed as
    \begin{align}
        \vxi_{\rm{2D}}(x,y)=\left\{ \begin{array}{ll} \frac{\mu_{12}}{2\pi \sigma^{2}}e^{-\frac{x^2 + y^2}{2\sigma^{2}}} \mathbf{e}_z, & \; \mathrm{if\ } -5 \mathrm{\ mu} \le x, y \le 5 \mathrm{\ mu}, \\ \\ 0, & \; \mathrm{otherwise}. \end{array} \right.
    \end{align}
    The dipole transition moment was set as $\mu_{12}=56.9$ Debye ($0.03$ mu), and the width of the Gaussian distribution was $\sigma = 4.771$ nm (1 mu). For spontaneous emission calculations, the initial electronic wave function was given as $\ket{\psi} = \sqrt{9/10}\ket{g} + \sqrt{1/10}\ket{e}$. Using such a small initial electronic excited-state population, as discussed around Eq. \eqref{eq:SE_sc_rate}, ensured a correct spontaneous emission rate as predicted by our semiclassical electrodynamics algorithm \cite{Li2018Spontaneous,Chen2018Spontaneous}. For all the simulations involving an external pulse, the TLS was initially set to the ground state. At each time step, the quantum dynamics of the TLS were propagated by the standard split-operator algorithm.

    \paragraph{Linear spectroscopy.} The linear spectroscopy in Figs. \ref{fig:setup}b,c were obtained by exciting the system with a plane wave. The source of the plane wave was centered at ($-2.13$ $\mu$m, 0) [($-447$ mu, 0)] and extended from $y = -996.74$ nm ($-208.92$ mu) to $y = 996.74$ nm ($208.92$ mu). Using such a wide source instead of a point source generated the plane wave in the spatial domain.  The time-dependent propagation of the plane wave was determined by a Gaussian source defined as $\left( -i\omega \right)^{-1} \partial_t \exp\left( -i\omega t - \frac{(t - t_0)^2}{2w^2} \right)$, with a center frequency $\omega = 4.1357$ eV (0.1 mu), a starting time $t_0= 0$, and an amplitude of $1$ mu. To generate Fig. \ref{fig:setup}b over a wide frequency range, the width of the  Gaussian source was set as $w = 3.3086$ eV (0.08 mu). In contrast, to generate the spectra in Fig. \ref{fig:setup}c over a narrower frequency window, the width of the Gaussian source was set to $1.6543$ eV ($0.04$ mu).

     The linear spectra were obtained by evaluating the flux spectrum using a simulation trajectory of approximately $1.8$ ps ($18000$ mu) with the standard MEEP routine.  The transmittance flux spectrum was computed by evaluating the power of the Poynting vector in the normal  direction ($\hat{\boldsymbol{\mathrm{n}}}$) as:
    \begin{align}
        P(\omega)=\mathrm{Re}\; \hat{\boldsymbol{\mathrm{n}}} \cdot \int \boldsymbol{\mathrm{E}}_\omega (x,y)^{\ast}\times \boldsymbol{\mathrm{H}}_{\omega}(x,y)\:dxdy .
    \end{align}
    Here, Re denotes the real component; $\boldsymbol{\mathrm{E}}_\omega$ and $\boldsymbol{\mathrm{H}}_{\omega}$ represent the complex-valued electric and magnetize fields in the frequency domain, respectively; and the asterisk symbol ($\ast$) denotes the complex conjugate. The spatial integration was performed within a 2D box  located at ($1.31$ $\mu$m, 0) [($275$ mu, 0)] and with dimensions $4.771\times4.771 \mathrm{\ nm}^2$ ($1.0\times1.0 \: \mathrm{mu}^2$).

    \paragraph{Spontaneous emission.} 
    For the spontaneous emission simulations shown in Figs. \ref{fig:setup}d, \ref{fig:Purcell_effect} and \ref{fig:radiative_decay_dependence}, the Gaussian pulse was deactivated. Each simulation trajectory was calculated for $10$ ps (100000 mu). The decay rate $k$ was then determined by fitting the population dynamics to the exponential function $\rho_{22}(t) = \rho_{22}(t=0)\;\mathrm{exp}\left( -kt \right)$.

    \paragraph{Driven polariton dynamics.} 
    For the results in Figs. \ref{fig:polariton_relaxation} and \ref{fig:TLS_Pulse}, in the time-domain, the Gaussian source width was set to $0.4136 $ eV (0.01 mu) to selectively excite a single polariton. Additionally, the amplitude of the Gaussian source was set to $10^{-3}$ mu, which was sufficiently weak to ensure that the TLS was not strongly excited. Each trajectory spanned  $4$ ps ($40000$ mu). The total EM energy was calculated using:
    \begin{align}
        U = \int \left( \frac{\varepsilon}{2} |\mathbf{E}(x,y)|^2 + \frac{1}{2\mu} |\mathbf{B}(x,y)|^2 \right) dxdy
    \end{align}
    integrated within the confined region between the two cavity mirrors. For the polariton energy dynamics shown in Fig. \ref{fig:polariton_relaxation}, the TLS was decoupled from the strong coupling system. For the time-dependent polariton energy trajectories, a moving average with a duration of 20 fs was applied to post-process the signals. The fitted polariton decay rates $k$ in Fig. \ref{fig:polariton_relaxation}c were obtained by performing an exponential fit $ y = a\mathrm{exp}\left( -kt \right) +b$ over the time window  [0.5 ps, 4 ps].
    
    \section{Linear spectroscopy versus $\mu_{12}$}

\begin{figure}[h]
	    \centering
	    \includegraphics[width=1.0\linewidth]{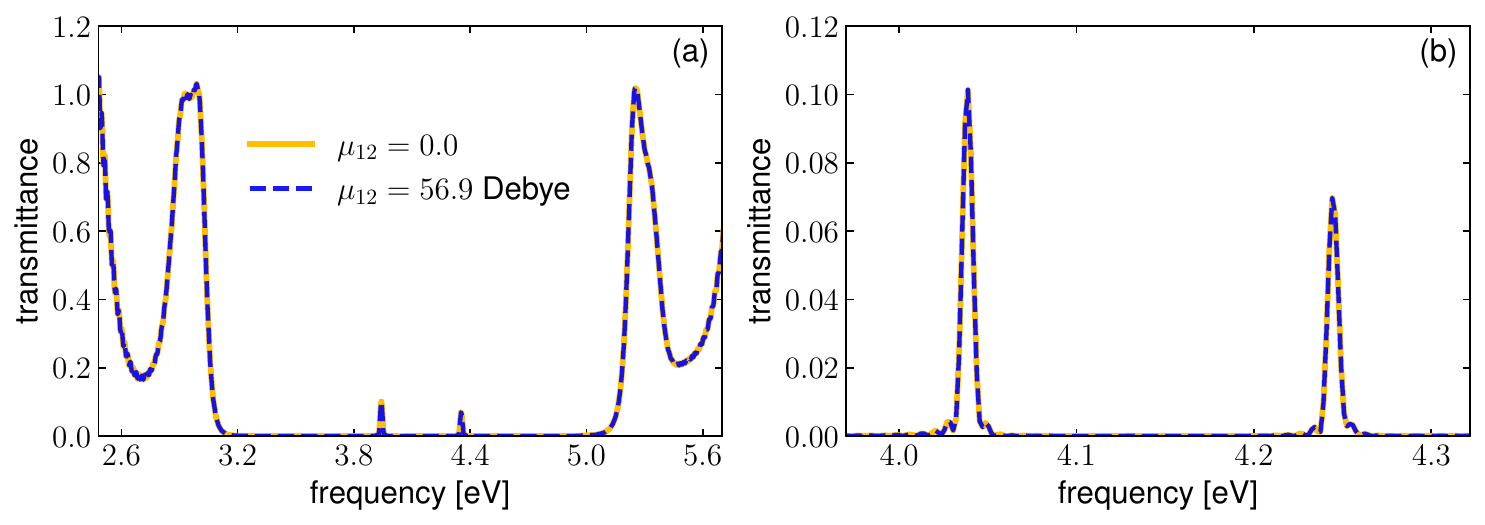}
	    \caption{The effects of the TLS on the linear polariton spectrum.  The spectra in the absence of the TLS (orange) in parts (a) and (b) are replotted from Figs. \ref{fig:setup}b,c with a Rabi splitting of 0.4127 eV. Including the TLS (dashed blue) does not affect the polariton spectrum.
        }
	    \label{fig:LinearSpectroscopy}
    \end{figure}

    As shown in Fig. \ref{fig:LinearSpectroscopy}, the presence of the TLS does not alter the linear polariton spectrum, suggesting that the TLS is only weakly coupled to the cavity.
    
    \setcounter{equation}{0}
	\renewcommand{\theequation}{E\arabic{equation}}
    
   \section{\red{The single-mode approximation}}\label{appendix:single-mode}
   \red{Throughout this manuscript, we have explicitly employed a 2D Bragg resonator to represent the cavity geometry. Here, we replace the 2D Bragg resonator with a single classical harmonic oscillator. Under this single-mode approximation, the semiclassical Hamiltonian for the TLS becomes:
   \begin{equation}
       \hH_{\rm {sc}} = \hH_{\rm s} - g_0 x_{\rm c} \hat{\sigma}_{x} .
   \end{equation}
   Here, $\hH_{\rm s}$ has been defined in Eq. \eqref{eq:LM_Hamiltonian}; $g_0$ represents the light-matter coupling constant; $x_{\rm c}$ denotes the position variable of the cavity mode; $\hat{\sigma}_{x} = \begin{pmatrix} 0  &  1  \\ 1 & 0  \end{pmatrix}$ is the Pauli $x$-matrix.
   
   The dynamics of the TLS is governed by the following quantum von Neumann equation:
   \begin{equation}
        \frac{\mathrm{d}}{\mathrm{d}t}\hat{\rho}(t)=-\frac{i}{\hbar}[\hat{H}_{\rm{sc}}(t),\hat{\rho}(t)] - \gamma_{\rm SE}
        \begin{pmatrix}
            -\rho_{\rm{ee}} & \frac{1}{2}\rho_{\rm{ge}} \\
            \frac{1}{2}\rho_{\rm{eg}} & \rho_{\rm{ee}}
        \end{pmatrix}
        . 
    \end{equation}
    Here, the last term represents the radiative dissipation of the TLS in vacuum, with $\gamma_{\rm SE}$ denoting the vacuum spontaneous emission rate of the TLS. By including this dissipation term, an explicit simulation of the electromagnetic environment is avoided.

    The cavity mode is expressed as a classical harmonic oscillator. The  equation of motion for the cavity mode reads
    \begin{subequations}
        \begin{align}
            \dot{p}_{\rm c} &=  - \omega_{\rm c}^2 q_{\rm c} + g_0 \tr{\hat{\rho} \hat{\sigma}_x} - \gamma_{\rm c} p_{\rm c} , \\
            \dot{x}_{\rm c} & = p_{\rm c}   ,
        \end{align}
    \end{subequations}
    where $\gamma_{\rm c}$ denotes the dissipation rate of the cavity mode. 

    \begin{figure}[h]
	    \centering
	    \includegraphics[width=1.0\linewidth]{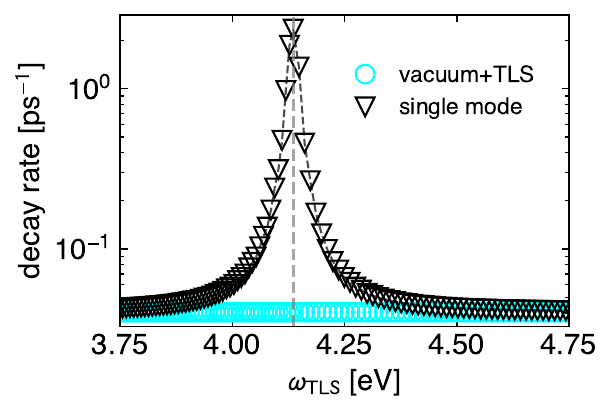}
	    \caption{\red{Fitted TLS decay rate as a function of $\omega_{\rm{TLS}}$, obtained using the single-mode model described in Appendix \ref{appendix:single-mode}. Two conditions are compared: inside the cavity (black triangles) versus in vacuum (cyan circles). The cavity frequency is indicated by the vertical dashed line. Simulation parameters are $g_0 = 10^{-4}$ mu, $\gamma_{\rm c} = 1.6\times 10^{-4}$ mu, and $\gamma_{\rm{SE}} = 4\times 10^{-6}$ mu.}
        }
	    \label{fig:single_mode}
    \end{figure}

    Fig. \ref{fig:single_mode} shows the fitted TLS decay rate under the single-mode approximation, analogous to the results in Fig. \ref{fig:Purcell_effect}. Within this single-mode approximation, the resonance dependence on the cavity mode frequency is well captured, consistent with Fig. \ref{fig:Purcell_effect}. However, this approach fails to reproduce the asymmetric behavior of the decay rate when the TLS frequency deviates from that of the cavity. Although simulations under strong coupling have not been performed here, we anticipate that the asymmetric behavior observed in Fig. \ref{fig:radiative_decay_dependence}a would similarly not be captured by the single-mode model.
    }

    

\begin{thebibliography}{95}%
\makeatletter
\providecommand \@ifxundefined [1]{%
 \@ifx{#1\undefined}
}%
\providecommand \@ifnum [1]{%
 \ifnum #1\expandafter \@firstoftwo
 \else \expandafter \@secondoftwo
 \fi
}%
\providecommand \@ifx [1]{%
 \ifx #1\expandafter \@firstoftwo
 \else \expandafter \@secondoftwo
 \fi
}%
\providecommand \natexlab [1]{#1}%
\providecommand \enquote  [1]{``#1''}%
\providecommand \bibnamefont  [1]{#1}%
\providecommand \bibfnamefont [1]{#1}%
\providecommand \citenamefont [1]{#1}%
\providecommand \href@noop [0]{\@secondoftwo}%
\providecommand \href [0]{\begingroup \@sanitize@url \@href}%
\providecommand \@href[1]{\@@startlink{#1}\@@href}%
\providecommand \@@href[1]{\endgroup#1\@@endlink}%
\providecommand \@sanitize@url [0]{\catcode `\\12\catcode `\$12\catcode `\&12\catcode `\#12\catcode `\^12\catcode `\_12\catcode `\%12\relax}%
\providecommand \@@startlink[1]{}%
\providecommand \@@endlink[0]{}%
\providecommand \url  [0]{\begingroup\@sanitize@url \@url }%
\providecommand \@url [1]{\endgroup\@href {#1}{\urlprefix }}%
\providecommand \urlprefix  [0]{URL }%
\providecommand \Eprint [0]{\href }%
\providecommand \doibase [0]{https://doi.org/}%
\providecommand \selectlanguage [0]{\@gobble}%
\providecommand \bibinfo  [0]{\@secondoftwo}%
\providecommand \bibfield  [0]{\@secondoftwo}%
\providecommand \translation [1]{[#1]}%
\providecommand \BibitemOpen [0]{}%
\providecommand \bibitemStop [0]{}%
\providecommand \bibitemNoStop [0]{.\EOS\space}%
\providecommand \EOS [0]{\spacefactor3000\relax}%
\providecommand \BibitemShut  [1]{\csname bibitem#1\endcsname}%
\let\auto@bib@innerbib\@empty
\bibitem [{\citenamefont {Hutchison}\ \emph {et~al.}(2012)\citenamefont {Hutchison}, \citenamefont {Schwartz}, \citenamefont {Genet}, \citenamefont {Devaux},\ and\ \citenamefont {Ebbesen}}]{Hutchison2012}%
  \BibitemOpen
  \bibfield  {author} {\bibinfo {author} {\bibfnamefont {J.~A.}\ \bibnamefont {Hutchison}}, \bibinfo {author} {\bibfnamefont {T.}~\bibnamefont {Schwartz}}, \bibinfo {author} {\bibfnamefont {C.}~\bibnamefont {Genet}}, \bibinfo {author} {\bibfnamefont {E.}~\bibnamefont {Devaux}},\ and\ \bibinfo {author} {\bibfnamefont {T.~W.}\ \bibnamefont {Ebbesen}},\ }\bibfield  {title} {\enquote {\bibinfo {title} {{Modifying Chemical Landscapes by Coupling to Vacuum Fields}},}\ }\href {https://doi.org/10.1002/anie.201107033} {\bibfield  {journal} {\bibinfo  {journal} {Angew. Chemie Int. Ed.}\ }\textbf {\bibinfo {volume} {51}},\ \bibinfo {pages} {1592--1596} (\bibinfo {year} {2012})}\BibitemShut {NoStop}%
\bibitem [{\citenamefont {Long}\ and\ \citenamefont {Simpkins}(2015)}]{Long2015}%
  \BibitemOpen
  \bibfield  {author} {\bibinfo {author} {\bibfnamefont {J.~P.}\ \bibnamefont {Long}}\ and\ \bibinfo {author} {\bibfnamefont {B.~S.}\ \bibnamefont {Simpkins}},\ }\bibfield  {title} {\enquote {\bibinfo {title} {{Coherent Coupling between a Molecular Vibration and Fabry–Perot Optical Cavity to Give Hybridized States in the Strong Coupling Limit}},}\ }\href {https://doi.org/10.1021/ph5003347} {\bibfield  {journal} {\bibinfo  {journal} {ACS Photonics}\ }\textbf {\bibinfo {volume} {2}},\ \bibinfo {pages} {130--136} (\bibinfo {year} {2015})}\BibitemShut {NoStop}%
\bibitem [{\citenamefont {Shalabney}\ \emph {et~al.}(2015{\natexlab{a}})\citenamefont {Shalabney}, \citenamefont {George}, \citenamefont {Hutchison}, \citenamefont {Pupillo}, \citenamefont {Genet},\ and\ \citenamefont {Ebbesen}}]{Shalabney2015}%
  \BibitemOpen
  \bibfield  {author} {\bibinfo {author} {\bibfnamefont {A.}~\bibnamefont {Shalabney}}, \bibinfo {author} {\bibfnamefont {J.}~\bibnamefont {George}}, \bibinfo {author} {\bibfnamefont {J.}~\bibnamefont {Hutchison}}, \bibinfo {author} {\bibfnamefont {G.}~\bibnamefont {Pupillo}}, \bibinfo {author} {\bibfnamefont {C.}~\bibnamefont {Genet}},\ and\ \bibinfo {author} {\bibfnamefont {T.~W.}\ \bibnamefont {Ebbesen}},\ }\bibfield  {title} {\enquote {\bibinfo {title} {{Coherent Coupling of Molecular Resonators with a Microcavity Mode}},}\ }\href {https://doi.org/10.1038/ncomms6981} {\bibfield  {journal} {\bibinfo  {journal} {Nat. Commun.}\ }\textbf {\bibinfo {volume} {6}},\ \bibinfo {pages} {5981} (\bibinfo {year} {2015}{\natexlab{a}})}\BibitemShut {NoStop}%
\bibitem [{\citenamefont {Dunkelberger}\ \emph {et~al.}(2016)\citenamefont {Dunkelberger}, \citenamefont {Spann}, \citenamefont {Fears}, \citenamefont {Simpkins},\ and\ \citenamefont {Owrutsky}}]{Dunkelberger2016}%
  \BibitemOpen
  \bibfield  {author} {\bibinfo {author} {\bibfnamefont {A.~D.}\ \bibnamefont {Dunkelberger}}, \bibinfo {author} {\bibfnamefont {B.~T.}\ \bibnamefont {Spann}}, \bibinfo {author} {\bibfnamefont {K.~P.}\ \bibnamefont {Fears}}, \bibinfo {author} {\bibfnamefont {B.~S.}\ \bibnamefont {Simpkins}},\ and\ \bibinfo {author} {\bibfnamefont {J.~C.}\ \bibnamefont {Owrutsky}},\ }\bibfield  {title} {\enquote {\bibinfo {title} {{Modified relaxation dynamics and coherent energy exchange in coupled vibration-cavity polaritons}},}\ }\href {https://doi.org/10.1038/ncomms13504} {\bibfield  {journal} {\bibinfo  {journal} {Nat. Commun.}\ }\textbf {\bibinfo {volume} {7}},\ \bibinfo {pages} {1--10} (\bibinfo {year} {2016})}\BibitemShut {NoStop}%
\bibitem [{\citenamefont {Ebbesen}, \citenamefont {Rubio},\ and\ \citenamefont {Scholes}(2023)}]{Ebbesen2023}%
  \BibitemOpen
  \bibfield  {author} {\bibinfo {author} {\bibfnamefont {T.~W.}\ \bibnamefont {Ebbesen}}, \bibinfo {author} {\bibfnamefont {A.}~\bibnamefont {Rubio}},\ and\ \bibinfo {author} {\bibfnamefont {G.~D.}\ \bibnamefont {Scholes}},\ }\bibfield  {title} {\enquote {\bibinfo {title} {{Introduction: Polaritonic Chemistry}},}\ }\href {https://doi.org/10.1021/ACS.CHEMREV.3C00637} {\bibfield  {journal} {\bibinfo  {journal} {Chem. Rev.}\ }\textbf {\bibinfo {volume} {123}},\ \bibinfo {pages} {12037--12038} (\bibinfo {year} {2023})}\BibitemShut {NoStop}%
\bibitem [{\citenamefont {Hirai}, \citenamefont {Hutchison},\ and\ \citenamefont {Uji-I}(2023)}]{Hirai2023}%
  \BibitemOpen
  \bibfield  {author} {\bibinfo {author} {\bibfnamefont {K.}~\bibnamefont {Hirai}}, \bibinfo {author} {\bibfnamefont {J.~A.}\ \bibnamefont {Hutchison}},\ and\ \bibinfo {author} {\bibfnamefont {H.}~\bibnamefont {Uji-I}},\ }\bibfield  {title} {\enquote {\bibinfo {title} {{Molecular Chemistry in Cavity Strong Coupling}},}\ }\href {https://doi.org/10.1021/ACS.CHEMREV.2C00748} {\bibfield  {journal} {\bibinfo  {journal} {Chem. Rev.}\ }\textbf {\bibinfo {volume} {123}},\ \bibinfo {pages} {8099--8126} (\bibinfo {year} {2023})}\BibitemShut {NoStop}%
\bibitem [{\citenamefont {Simpkins}, \citenamefont {Dunkelberger},\ and\ \citenamefont {Vurgaftman}(2023)}]{Simpkins2023}%
  \BibitemOpen
  \bibfield  {author} {\bibinfo {author} {\bibfnamefont {B.~S.}\ \bibnamefont {Simpkins}}, \bibinfo {author} {\bibfnamefont {A.~D.}\ \bibnamefont {Dunkelberger}},\ and\ \bibinfo {author} {\bibfnamefont {I.}~\bibnamefont {Vurgaftman}},\ }\bibfield  {title} {\enquote {\bibinfo {title} {{Control, Modulation, and Analytical Descriptions of Vibrational Strong Coupling}},}\ }\href {https://doi.org/10.1021/acs.chemrev.2c00774} {\bibfield  {journal} {\bibinfo  {journal} {Chem. Rev.}\ }\textbf {\bibinfo {volume} {123}},\ \bibinfo {pages} {5020--5048} (\bibinfo {year} {2023})}\BibitemShut {NoStop}%
\bibitem [{\citenamefont {Mandal}\ \emph {et~al.}(2023)\citenamefont {Mandal}, \citenamefont {Taylor}, \citenamefont {Weight}, \citenamefont {Koessler}, \citenamefont {Li},\ and\ \citenamefont {Huo}}]{Mandal2023ChemRev}%
  \BibitemOpen
  \bibfield  {author} {\bibinfo {author} {\bibfnamefont {A.}~\bibnamefont {Mandal}}, \bibinfo {author} {\bibfnamefont {M.~A.}\ \bibnamefont {Taylor}}, \bibinfo {author} {\bibfnamefont {B.~M.}\ \bibnamefont {Weight}}, \bibinfo {author} {\bibfnamefont {E.~R.}\ \bibnamefont {Koessler}}, \bibinfo {author} {\bibfnamefont {X.}~\bibnamefont {Li}},\ and\ \bibinfo {author} {\bibfnamefont {P.}~\bibnamefont {Huo}},\ }\bibfield  {title} {\enquote {\bibinfo {title} {{Theoretical Advances in Polariton Chemistry and Molecular Cavity Quantum Electrodynamics}},}\ }\href {https://doi.org/10.1021/acs.chemrev.2c00855} {\bibfield  {journal} {\bibinfo  {journal} {Chem. Rev.}\ }\textbf {\bibinfo {volume} {123}},\ \bibinfo {pages} {9786--9879} (\bibinfo {year} {2023})}\BibitemShut {NoStop}%
\bibitem [{\citenamefont {Bhuyan}\ \emph {et~al.}(2023)\citenamefont {Bhuyan}, \citenamefont {Mony}, \citenamefont {Kotov}, \citenamefont {Castellanos}, \citenamefont {{G{\'{o}}mez Rivas}}, \citenamefont {Shegai},\ and\ \citenamefont {B{\"{o}}rjesson}}]{Bhuyan2023}%
  \BibitemOpen
  \bibfield  {author} {\bibinfo {author} {\bibfnamefont {R.}~\bibnamefont {Bhuyan}}, \bibinfo {author} {\bibfnamefont {J.}~\bibnamefont {Mony}}, \bibinfo {author} {\bibfnamefont {O.}~\bibnamefont {Kotov}}, \bibinfo {author} {\bibfnamefont {G.~W.}\ \bibnamefont {Castellanos}}, \bibinfo {author} {\bibfnamefont {J.}~\bibnamefont {{G{\'{o}}mez Rivas}}}, \bibinfo {author} {\bibfnamefont {T.~O.}\ \bibnamefont {Shegai}},\ and\ \bibinfo {author} {\bibfnamefont {K.}~\bibnamefont {B{\"{o}}rjesson}},\ }\bibfield  {title} {\enquote {\bibinfo {title} {{The Rise and Current Status of Polaritonic Photochemistry and Photophysics}},}\ }\href {https://doi.org/10.1021/ACS.CHEMREV.2C00895} {\bibfield  {journal} {\bibinfo  {journal} {Chem. Rev.}\ }\textbf {\bibinfo {volume} {123}},\ \bibinfo {pages} {10877--10919} (\bibinfo {year} {2023})}\BibitemShut {NoStop}%
\bibitem [{\citenamefont {Ruggenthaler}, \citenamefont {Sidler},\ and\ \citenamefont {Rubio}(2023)}]{Ruggenthaler2023}%
  \BibitemOpen
  \bibfield  {author} {\bibinfo {author} {\bibfnamefont {M.}~\bibnamefont {Ruggenthaler}}, \bibinfo {author} {\bibfnamefont {D.}~\bibnamefont {Sidler}},\ and\ \bibinfo {author} {\bibfnamefont {A.}~\bibnamefont {Rubio}},\ }\bibfield  {title} {\enquote {\bibinfo {title} {{Understanding Polaritonic Chemistry from Ab Initio Quantum Electrodynamics}},}\ }\href {https://doi.org/10.1021/acs.chemrev.2c00788} {\bibfield  {journal} {\bibinfo  {journal} {Chem. Rev.}\ }\textbf {\bibinfo {volume} {123}},\ \bibinfo {pages} {11191--11229} (\bibinfo {year} {2023})}\BibitemShut {NoStop}%
\bibitem [{\citenamefont {Tibben}\ \emph {et~al.}(2023)\citenamefont {Tibben}, \citenamefont {Bonin}, \citenamefont {Cho}, \citenamefont {Lakhwani}, \citenamefont {Hutchison},\ and\ \citenamefont {G{\'{o}}mez}}]{Tibben2023}%
  \BibitemOpen
  \bibfield  {author} {\bibinfo {author} {\bibfnamefont {D.~J.}\ \bibnamefont {Tibben}}, \bibinfo {author} {\bibfnamefont {G.~O.}\ \bibnamefont {Bonin}}, \bibinfo {author} {\bibfnamefont {I.}~\bibnamefont {Cho}}, \bibinfo {author} {\bibfnamefont {G.}~\bibnamefont {Lakhwani}}, \bibinfo {author} {\bibfnamefont {J.}~\bibnamefont {Hutchison}},\ and\ \bibinfo {author} {\bibfnamefont {D.~E.}\ \bibnamefont {G{\'{o}}mez}},\ }\bibfield  {title} {\enquote {\bibinfo {title} {{Molecular Energy Transfer under the Strong Light-Matter Interaction Regime}},}\ }\href {https://doi.org/10.1021/ACS.CHEMREV.2C00702} {\bibfield  {journal} {\bibinfo  {journal} {Chem. Rev.}\ }\textbf {\bibinfo {volume} {123}},\ \bibinfo {pages} {8044--8068} (\bibinfo {year} {2023})}\BibitemShut {NoStop}%
\bibitem [{\citenamefont {Xiang}\ and\ \citenamefont {Xiong}(2024)}]{Xiang2024}%
  \BibitemOpen
  \bibfield  {author} {\bibinfo {author} {\bibfnamefont {B.}~\bibnamefont {Xiang}}\ and\ \bibinfo {author} {\bibfnamefont {W.}~\bibnamefont {Xiong}},\ }\bibfield  {title} {\enquote {\bibinfo {title} {{Molecular Polaritons for Chemistry, Photonics and Quantum Technologies}},}\ }\href {https://doi.org/10.1021/ACS.CHEMREV.3C00662} {\bibfield  {journal} {\bibinfo  {journal} {Chem. Rev.}\ }\textbf {\bibinfo {volume} {124}},\ \bibinfo {pages} {2512--2552} (\bibinfo {year} {2024})}\BibitemShut {NoStop}%
\bibitem [{\citenamefont {Zhong}\ \emph {et~al.}(2017)\citenamefont {Zhong}, \citenamefont {Chervy}, \citenamefont {Zhang}, \citenamefont {Thomas}, \citenamefont {George}, \citenamefont {Genet}, \citenamefont {Hutchison},\ and\ \citenamefont {Ebbesen}}]{Zhong2017}%
  \BibitemOpen
  \bibfield  {author} {\bibinfo {author} {\bibfnamefont {X.}~\bibnamefont {Zhong}}, \bibinfo {author} {\bibfnamefont {T.}~\bibnamefont {Chervy}}, \bibinfo {author} {\bibfnamefont {L.}~\bibnamefont {Zhang}}, \bibinfo {author} {\bibfnamefont {A.}~\bibnamefont {Thomas}}, \bibinfo {author} {\bibfnamefont {J.}~\bibnamefont {George}}, \bibinfo {author} {\bibfnamefont {C.}~\bibnamefont {Genet}}, \bibinfo {author} {\bibfnamefont {J.~A.}\ \bibnamefont {Hutchison}},\ and\ \bibinfo {author} {\bibfnamefont {T.~W.}\ \bibnamefont {Ebbesen}},\ }\bibfield  {title} {\enquote {\bibinfo {title} {{Energy Transfer between Spatially Separated Entangled Molecules}},}\ }\href {https://doi.org/10.1002/anie.201703539} {\bibfield  {journal} {\bibinfo  {journal} {Angew. Chemie Int. Ed.}\ }\textbf {\bibinfo {volume} {56}},\ \bibinfo {pages} {9034--9038} (\bibinfo {year} {2017})}\BibitemShut {NoStop}%
\bibitem [{\citenamefont {Xiang}\ \emph {et~al.}(2020)\citenamefont {Xiang}, \citenamefont {Ribeiro}, \citenamefont {Du}, \citenamefont {Chen}, \citenamefont {Yang}, \citenamefont {Wang}, \citenamefont {Yuen-Zhou},\ and\ \citenamefont {Xiong}}]{Xiang2020Science}%
  \BibitemOpen
  \bibfield  {author} {\bibinfo {author} {\bibfnamefont {B.}~\bibnamefont {Xiang}}, \bibinfo {author} {\bibfnamefont {R.~F.}\ \bibnamefont {Ribeiro}}, \bibinfo {author} {\bibfnamefont {M.}~\bibnamefont {Du}}, \bibinfo {author} {\bibfnamefont {L.}~\bibnamefont {Chen}}, \bibinfo {author} {\bibfnamefont {Z.}~\bibnamefont {Yang}}, \bibinfo {author} {\bibfnamefont {J.}~\bibnamefont {Wang}}, \bibinfo {author} {\bibfnamefont {J.}~\bibnamefont {Yuen-Zhou}},\ and\ \bibinfo {author} {\bibfnamefont {W.}~\bibnamefont {Xiong}},\ }\bibfield  {title} {\enquote {\bibinfo {title} {{Intermolecular Vibrational Energy Transfer Enabled by Microcavity Strong Light--Matter Coupling}},}\ }\href {https://doi.org/10.1126/science.aba3544} {\bibfield  {journal} {\bibinfo  {journal} {Science}\ }\textbf {\bibinfo {volume} {368}},\ \bibinfo {pages} {665--667} (\bibinfo {year} {2020})}\BibitemShut {NoStop}%
\bibitem [{\citenamefont {Chen}\ \emph {et~al.}(2022)\citenamefont {Chen}, \citenamefont {Du}, \citenamefont {Yang}, \citenamefont {Yuen-Zhou},\ and\ \citenamefont {Xiong}}]{Chen2022}%
  \BibitemOpen
  \bibfield  {author} {\bibinfo {author} {\bibfnamefont {T.-T.}\ \bibnamefont {Chen}}, \bibinfo {author} {\bibfnamefont {M.}~\bibnamefont {Du}}, \bibinfo {author} {\bibfnamefont {Z.}~\bibnamefont {Yang}}, \bibinfo {author} {\bibfnamefont {J.}~\bibnamefont {Yuen-Zhou}},\ and\ \bibinfo {author} {\bibfnamefont {W.}~\bibnamefont {Xiong}},\ }\bibfield  {title} {\enquote {\bibinfo {title} {{Cavity-enabled Enhancement of Ultrafast Intramolecular Vibrational Redistribution over Pseudorotation}},}\ }\href {https://doi.org/10.1126/science.add0276} {\bibfield  {journal} {\bibinfo  {journal} {Science}\ }\textbf {\bibinfo {volume} {378}},\ \bibinfo {pages} {790--794} (\bibinfo {year} {2022})}\BibitemShut {NoStop}%
\bibitem [{\citenamefont {Thomas}\ \emph {et~al.}(2016)\citenamefont {Thomas}, \citenamefont {George}, \citenamefont {Shalabney}, \citenamefont {Dryzhakov}, \citenamefont {Varma}, \citenamefont {Moran}, \citenamefont {Chervy}, \citenamefont {Zhong}, \citenamefont {Devaux}, \citenamefont {Genet}, \citenamefont {Hutchison},\ and\ \citenamefont {Ebbesen}}]{Thomas2016}%
  \BibitemOpen
  \bibfield  {author} {\bibinfo {author} {\bibfnamefont {A.}~\bibnamefont {Thomas}}, \bibinfo {author} {\bibfnamefont {J.}~\bibnamefont {George}}, \bibinfo {author} {\bibfnamefont {A.}~\bibnamefont {Shalabney}}, \bibinfo {author} {\bibfnamefont {M.}~\bibnamefont {Dryzhakov}}, \bibinfo {author} {\bibfnamefont {S.~J.}\ \bibnamefont {Varma}}, \bibinfo {author} {\bibfnamefont {J.}~\bibnamefont {Moran}}, \bibinfo {author} {\bibfnamefont {T.}~\bibnamefont {Chervy}}, \bibinfo {author} {\bibfnamefont {X.}~\bibnamefont {Zhong}}, \bibinfo {author} {\bibfnamefont {E.}~\bibnamefont {Devaux}}, \bibinfo {author} {\bibfnamefont {C.}~\bibnamefont {Genet}}, \bibinfo {author} {\bibfnamefont {J.~A.}\ \bibnamefont {Hutchison}},\ and\ \bibinfo {author} {\bibfnamefont {T.~W.}\ \bibnamefont {Ebbesen}},\ }\bibfield  {title} {\enquote {\bibinfo {title} {{Ground-State Chemical Reactivity under Vibrational Coupling to the Vacuum Electromagnetic Field}},}\ }\href {https://doi.org/10.1002/anie.201605504} {\bibfield  {journal} {\bibinfo
  {journal} {Angew. Chemie Int. Ed.}\ }\textbf {\bibinfo {volume} {55}},\ \bibinfo {pages} {11462--11466} (\bibinfo {year} {2016})}\BibitemShut {NoStop}%
\bibitem [{\citenamefont {Thomas}\ \emph {et~al.}(2019)\citenamefont {Thomas}, \citenamefont {Lethuillier-Karl}, \citenamefont {Nagarajan}, \citenamefont {Vergauwe}, \citenamefont {George}, \citenamefont {Chervy}, \citenamefont {Shalabney}, \citenamefont {Devaux}, \citenamefont {Genet}, \citenamefont {Moran},\ and\ \citenamefont {Ebbesen}}]{Thomas2019_science}%
  \BibitemOpen
  \bibfield  {author} {\bibinfo {author} {\bibfnamefont {A.}~\bibnamefont {Thomas}}, \bibinfo {author} {\bibfnamefont {L.}~\bibnamefont {Lethuillier-Karl}}, \bibinfo {author} {\bibfnamefont {K.}~\bibnamefont {Nagarajan}}, \bibinfo {author} {\bibfnamefont {R.~M.~A.}\ \bibnamefont {Vergauwe}}, \bibinfo {author} {\bibfnamefont {J.}~\bibnamefont {George}}, \bibinfo {author} {\bibfnamefont {T.}~\bibnamefont {Chervy}}, \bibinfo {author} {\bibfnamefont {A.}~\bibnamefont {Shalabney}}, \bibinfo {author} {\bibfnamefont {E.}~\bibnamefont {Devaux}}, \bibinfo {author} {\bibfnamefont {C.}~\bibnamefont {Genet}}, \bibinfo {author} {\bibfnamefont {J.}~\bibnamefont {Moran}},\ and\ \bibinfo {author} {\bibfnamefont {T.~W.}\ \bibnamefont {Ebbesen}},\ }\bibfield  {title} {\enquote {\bibinfo {title} {{Tilting a Ground-State Reactivity Landscape by Vibrational Strong Coupling}},}\ }\href {https://doi.org/10.1126/science.aau7742} {\bibfield  {journal} {\bibinfo  {journal} {Science}\ }\textbf {\bibinfo {volume} {363}},\ \bibinfo
  {pages} {615--619} (\bibinfo {year} {2019})}\BibitemShut {NoStop}%
\bibitem [{\citenamefont {Ahn}\ \emph {et~al.}(2023)\citenamefont {Ahn}, \citenamefont {Triana}, \citenamefont {Recabal}, \citenamefont {Herrera},\ and\ \citenamefont {Simpkins}}]{Ahn2023Science}%
  \BibitemOpen
  \bibfield  {author} {\bibinfo {author} {\bibfnamefont {W.}~\bibnamefont {Ahn}}, \bibinfo {author} {\bibfnamefont {J.~F.}\ \bibnamefont {Triana}}, \bibinfo {author} {\bibfnamefont {F.}~\bibnamefont {Recabal}}, \bibinfo {author} {\bibfnamefont {F.}~\bibnamefont {Herrera}},\ and\ \bibinfo {author} {\bibfnamefont {B.~S.}\ \bibnamefont {Simpkins}},\ }\bibfield  {title} {\enquote {\bibinfo {title} {{Modification of ground-state chemical reactivity via light–matter coherence in infrared cavities}},}\ }\href {https://doi.org/10.1126/science.ade7147} {\bibfield  {journal} {\bibinfo  {journal} {Science}\ }\textbf {\bibinfo {volume} {380}},\ \bibinfo {pages} {1165--1168} (\bibinfo {year} {2023})}\BibitemShut {NoStop}%
\bibitem [{\citenamefont {Chikkaraddy}\ \emph {et~al.}(2016)\citenamefont {Chikkaraddy}, \citenamefont {de~Nijs}, \citenamefont {Benz}, \citenamefont {Barrow}, \citenamefont {Scherman}, \citenamefont {Rosta}, \citenamefont {Demetriadou}, \citenamefont {Fox}, \citenamefont {Hess},\ and\ \citenamefont {Baumberg}}]{Chikkaraddy2016}%
  \BibitemOpen
  \bibfield  {author} {\bibinfo {author} {\bibfnamefont {R.}~\bibnamefont {Chikkaraddy}}, \bibinfo {author} {\bibfnamefont {B.}~\bibnamefont {de~Nijs}}, \bibinfo {author} {\bibfnamefont {F.}~\bibnamefont {Benz}}, \bibinfo {author} {\bibfnamefont {S.~J.}\ \bibnamefont {Barrow}}, \bibinfo {author} {\bibfnamefont {O.~A.}\ \bibnamefont {Scherman}}, \bibinfo {author} {\bibfnamefont {E.}~\bibnamefont {Rosta}}, \bibinfo {author} {\bibfnamefont {A.}~\bibnamefont {Demetriadou}}, \bibinfo {author} {\bibfnamefont {P.}~\bibnamefont {Fox}}, \bibinfo {author} {\bibfnamefont {O.}~\bibnamefont {Hess}},\ and\ \bibinfo {author} {\bibfnamefont {J.~J.}\ \bibnamefont {Baumberg}},\ }\bibfield  {title} {\enquote {\bibinfo {title} {{Single-molecule strong coupling at room temperature in plasmonic nanocavities}},}\ }\href {https://doi.org/10.1038/nature17974} {\bibfield  {journal} {\bibinfo  {journal} {Nature}\ }\textbf {\bibinfo {volume} {535}},\ \bibinfo {pages} {127--130} (\bibinfo {year} {2016})}\BibitemShut {NoStop}%
\bibitem [{\citenamefont {Yoo}\ \emph {et~al.}(2021)\citenamefont {Yoo}, \citenamefont {de~Le{\'{o}}n-P{\'{e}}rez}, \citenamefont {Pelton}, \citenamefont {Lee}, \citenamefont {Mohr}, \citenamefont {Raschke}, \citenamefont {Caldwell}, \citenamefont {Mart{\'{i}}n-Moreno},\ and\ \citenamefont {Oh}}]{Yoo2021}%
  \BibitemOpen
  \bibfield  {author} {\bibinfo {author} {\bibfnamefont {D.}~\bibnamefont {Yoo}}, \bibinfo {author} {\bibfnamefont {F.}~\bibnamefont {de~Le{\'{o}}n-P{\'{e}}rez}}, \bibinfo {author} {\bibfnamefont {M.}~\bibnamefont {Pelton}}, \bibinfo {author} {\bibfnamefont {I.-H.}\ \bibnamefont {Lee}}, \bibinfo {author} {\bibfnamefont {D.~A.}\ \bibnamefont {Mohr}}, \bibinfo {author} {\bibfnamefont {M.~B.}\ \bibnamefont {Raschke}}, \bibinfo {author} {\bibfnamefont {J.~D.}\ \bibnamefont {Caldwell}}, \bibinfo {author} {\bibfnamefont {L.}~\bibnamefont {Mart{\'{i}}n-Moreno}},\ and\ \bibinfo {author} {\bibfnamefont {S.-H.}\ \bibnamefont {Oh}},\ }\bibfield  {title} {\enquote {\bibinfo {title} {{Ultrastrong plasmon–phonon coupling via epsilon-near-zero nanocavities}},}\ }\href {https://doi.org/10.1038/s41566-020-00731-5} {\bibfield  {journal} {\bibinfo  {journal} {Nat. Photonics}\ }\textbf {\bibinfo {volume} {15}},\ \bibinfo {pages} {125--130} (\bibinfo {year} {2021})}\BibitemShut {NoStop}%
\bibitem [{\citenamefont {Brawley}\ \emph {et~al.}(2021)\citenamefont {Brawley}, \citenamefont {Storm}, \citenamefont {{Contreras Mora}}, \citenamefont {Pelton},\ and\ \citenamefont {Sheldon}}]{Brawley2021}%
  \BibitemOpen
  \bibfield  {author} {\bibinfo {author} {\bibfnamefont {Z.~T.}\ \bibnamefont {Brawley}}, \bibinfo {author} {\bibfnamefont {S.~D.}\ \bibnamefont {Storm}}, \bibinfo {author} {\bibfnamefont {D.~A.}\ \bibnamefont {{Contreras Mora}}}, \bibinfo {author} {\bibfnamefont {M.}~\bibnamefont {Pelton}},\ and\ \bibinfo {author} {\bibfnamefont {M.}~\bibnamefont {Sheldon}},\ }\bibfield  {title} {\enquote {\bibinfo {title} {{Angle-independent plasmonic substrates for multi-mode vibrational strong coupling with molecular thin films}},}\ }\href {https://doi.org/10.1063/5.0039195} {\bibfield  {journal} {\bibinfo  {journal} {J. Chem. Phys.}\ }\textbf {\bibinfo {volume} {154}},\ \bibinfo {pages} {104305} (\bibinfo {year} {2021})}\BibitemShut {NoStop}%
\bibitem [{\citenamefont {Wright}, \citenamefont {Nelson},\ and\ \citenamefont {Weichman}(2023)}]{Wright2023}%
  \BibitemOpen
  \bibfield  {author} {\bibinfo {author} {\bibfnamefont {A.~D.}\ \bibnamefont {Wright}}, \bibinfo {author} {\bibfnamefont {J.~C.}\ \bibnamefont {Nelson}},\ and\ \bibinfo {author} {\bibfnamefont {M.~L.}\ \bibnamefont {Weichman}},\ }\bibfield  {title} {\enquote {\bibinfo {title} {{Rovibrational Polaritons in Gas-Phase Methane}},}\ }\href {https://doi.org/10.1021/jacs.3c00126} {\bibfield  {journal} {\bibinfo  {journal} {J. Am. Chem. Soc.}\ }\textbf {\bibinfo {volume} {145}},\ \bibinfo {pages} {5982--5987} (\bibinfo {year} {2023})}\BibitemShut {NoStop}%
\bibitem [{\citenamefont {Canales}\ \emph {et~al.}(2021)\citenamefont {Canales}, \citenamefont {Baranov}, \citenamefont {Antosiewicz},\ and\ \citenamefont {Shegai}}]{Canales2021}%
  \BibitemOpen
  \bibfield  {author} {\bibinfo {author} {\bibfnamefont {A.}~\bibnamefont {Canales}}, \bibinfo {author} {\bibfnamefont {D.~G.}\ \bibnamefont {Baranov}}, \bibinfo {author} {\bibfnamefont {T.~J.}\ \bibnamefont {Antosiewicz}},\ and\ \bibinfo {author} {\bibfnamefont {T.}~\bibnamefont {Shegai}},\ }\bibfield  {title} {\enquote {\bibinfo {title} {{Abundance of cavity-free polaritonic states in resonant materials and nanostructures}},}\ }\href {https://doi.org/10.1063/5.0033352} {\bibfield  {journal} {\bibinfo  {journal} {J. Chem. Phys.}\ }\textbf {\bibinfo {volume} {154}},\ \bibinfo {pages} {024701} (\bibinfo {year} {2021})}\BibitemShut {NoStop}%
\bibitem [{\citenamefont {Canales}\ \emph {et~al.}(2024)\citenamefont {Canales}, \citenamefont {Kotov}, \citenamefont {K{\"{u}}{\c{c}}{\"{u}}k{\"{o}}z},\ and\ \citenamefont {Shegai}}]{Canales2024}%
  \BibitemOpen
  \bibfield  {author} {\bibinfo {author} {\bibfnamefont {A.}~\bibnamefont {Canales}}, \bibinfo {author} {\bibfnamefont {O.~V.}\ \bibnamefont {Kotov}}, \bibinfo {author} {\bibfnamefont {B.}~\bibnamefont {K{\"{u}}{\c{c}}{\"{u}}k{\"{o}}z}},\ and\ \bibinfo {author} {\bibfnamefont {T.~O.}\ \bibnamefont {Shegai}},\ }\bibfield  {title} {\enquote {\bibinfo {title} {{Self-Hybridized Vibrational-Mie Polaritons in Water Droplets}},}\ }\href {https://doi.org/10.1103/PHYSREVLETT.132.193804} {\bibfield  {journal} {\bibinfo  {journal} {Phys. Rev. Lett.}\ }\textbf {\bibinfo {volume} {132}},\ \bibinfo {pages} {193804} (\bibinfo {year} {2024})}\BibitemShut {NoStop}%
\bibitem [{\citenamefont {Zhao}\ \emph {et~al.}(2024)\citenamefont {Zhao}, \citenamefont {Arneson}, \citenamefont {Fan},\ and\ \citenamefont {Forrest}}]{Zhao2024}%
  \BibitemOpen
  \bibfield  {author} {\bibinfo {author} {\bibfnamefont {H.}~\bibnamefont {Zhao}}, \bibinfo {author} {\bibfnamefont {C.~E.}\ \bibnamefont {Arneson}}, \bibinfo {author} {\bibfnamefont {D.}~\bibnamefont {Fan}},\ and\ \bibinfo {author} {\bibfnamefont {S.~R.}\ \bibnamefont {Forrest}},\ }\bibfield  {title} {\enquote {\bibinfo {title} {{Stable blue phosphorescent organic LEDs that use polariton-enhanced Purcell effects}},}\ }\href {https://doi.org/10.1038/s41586-023-06976-8} {\bibfield  {journal} {\bibinfo  {journal} {Nature}\ }\textbf {\bibinfo {volume} {626}},\ \bibinfo {pages} {300--305} (\bibinfo {year} {2024})}\BibitemShut {NoStop}%
\bibitem [{\citenamefont {Shalabney}\ \emph {et~al.}(2015{\natexlab{b}})\citenamefont {Shalabney}, \citenamefont {George}, \citenamefont {Hiura}, \citenamefont {Hutchison}, \citenamefont {Genet}, \citenamefont {Hellwig},\ and\ \citenamefont {Ebbesen}}]{Shalabney2015Raman}%
  \BibitemOpen
  \bibfield  {author} {\bibinfo {author} {\bibfnamefont {A.}~\bibnamefont {Shalabney}}, \bibinfo {author} {\bibfnamefont {J.}~\bibnamefont {George}}, \bibinfo {author} {\bibfnamefont {H.}~\bibnamefont {Hiura}}, \bibinfo {author} {\bibfnamefont {J.~A.}\ \bibnamefont {Hutchison}}, \bibinfo {author} {\bibfnamefont {C.}~\bibnamefont {Genet}}, \bibinfo {author} {\bibfnamefont {P.}~\bibnamefont {Hellwig}},\ and\ \bibinfo {author} {\bibfnamefont {T.~W.}\ \bibnamefont {Ebbesen}},\ }\bibfield  {title} {\enquote {\bibinfo {title} {{Enhanced Raman Scattering from Vibro-Polariton Hybrid States}},}\ }\href {https://doi.org/10.1002/ange.201502979} {\bibfield  {journal} {\bibinfo  {journal} {Angew. Chemie}\ }\textbf {\bibinfo {volume} {127}},\ \bibinfo {pages} {8082--8086} (\bibinfo {year} {2015}{\natexlab{b}})}\BibitemShut {NoStop}%
\bibitem [{\citenamefont {del Pino}, \citenamefont {Feist},\ and\ \citenamefont {Garcia-Vidal}(2015)}]{DelPino2015}%
  \BibitemOpen
  \bibfield  {author} {\bibinfo {author} {\bibfnamefont {J.}~\bibnamefont {del Pino}}, \bibinfo {author} {\bibfnamefont {J.}~\bibnamefont {Feist}},\ and\ \bibinfo {author} {\bibfnamefont {F.~J.}\ \bibnamefont {Garcia-Vidal}},\ }\bibfield  {title} {\enquote {\bibinfo {title} {{Signatures of Vibrational Strong Coupling in Raman Scattering}},}\ }\href {https://doi.org/10.1021/acs.jpcc.5b11654} {\bibfield  {journal} {\bibinfo  {journal} {J. Phys. Chem. C}\ }\textbf {\bibinfo {volume} {119}},\ \bibinfo {pages} {29132--29137} (\bibinfo {year} {2015})}\BibitemShut {NoStop}%
\bibitem [{\citenamefont {Ahn}\ and\ \citenamefont {Simpkins}(2021)}]{Ahn2021}%
  \BibitemOpen
  \bibfield  {author} {\bibinfo {author} {\bibfnamefont {W.}~\bibnamefont {Ahn}}\ and\ \bibinfo {author} {\bibfnamefont {B.~S.}\ \bibnamefont {Simpkins}},\ }\bibfield  {title} {\enquote {\bibinfo {title} {{Raman Scattering under Strong Vibration-Cavity Coupling}},}\ }\href {https://doi.org/10.1021/acs.jpcc.0c10360} {\bibfield  {journal} {\bibinfo  {journal} {J. Phys. Chem. C}\ }\textbf {\bibinfo {volume} {125}},\ \bibinfo {pages} {830--835} (\bibinfo {year} {2021})}\BibitemShut {NoStop}%
\bibitem [{\citenamefont {Nagarajan}, \citenamefont {Thomas},\ and\ \citenamefont {Ebbesen}(2021)}]{Nagarajan2021}%
  \BibitemOpen
  \bibfield  {author} {\bibinfo {author} {\bibfnamefont {K.}~\bibnamefont {Nagarajan}}, \bibinfo {author} {\bibfnamefont {A.}~\bibnamefont {Thomas}},\ and\ \bibinfo {author} {\bibfnamefont {T.~W.}\ \bibnamefont {Ebbesen}},\ }\bibfield  {title} {\enquote {\bibinfo {title} {{Chemistry under Vibrational Strong Coupling}},}\ }\href {https://doi.org/10.1021/jacs.1c07420} {\bibfield  {journal} {\bibinfo  {journal} {J. Am. Chem. Soc.}\ }\textbf {\bibinfo {volume} {143}},\ \bibinfo {pages} {16877--16889} (\bibinfo {year} {2021})}\BibitemShut {NoStop}%
\bibitem [{\citenamefont {Takele}\ \emph {et~al.}(2020)\citenamefont {Takele}, \citenamefont {Wackenhut}, \citenamefont {Piatkowski}, \citenamefont {Meixner},\ and\ \citenamefont {Waluk}}]{Takele2020}%
  \BibitemOpen
  \bibfield  {author} {\bibinfo {author} {\bibfnamefont {W.~M.}\ \bibnamefont {Takele}}, \bibinfo {author} {\bibfnamefont {F.}~\bibnamefont {Wackenhut}}, \bibinfo {author} {\bibfnamefont {L.}~\bibnamefont {Piatkowski}}, \bibinfo {author} {\bibfnamefont {A.~J.}\ \bibnamefont {Meixner}},\ and\ \bibinfo {author} {\bibfnamefont {J.}~\bibnamefont {Waluk}},\ }\bibfield  {title} {\enquote {\bibinfo {title} {{Multimode Vibrational Strong Coupling of Methyl Salicylate to a Fabry–P{\'{e}}rot Microcavity}},}\ }\href {https://doi.org/10.1021/acs.jpcb.0c03815} {\bibfield  {journal} {\bibinfo  {journal} {J. Phys. Chem. B}\ }\textbf {\bibinfo {volume} {124}},\ \bibinfo {pages} {5709--5716} (\bibinfo {year} {2020})}\BibitemShut {NoStop}%
\bibitem [{\citenamefont {Thomas}\ and\ \citenamefont {Barnes}(2024)}]{Thomas2024shifts}%
  \BibitemOpen
  \bibfield  {author} {\bibinfo {author} {\bibfnamefont {P.~A.}\ \bibnamefont {Thomas}}\ and\ \bibinfo {author} {\bibfnamefont {W.~L.}\ \bibnamefont {Barnes}},\ }\bibfield  {title} {\enquote {\bibinfo {title} {{Strong coupling-induced frequency shifts of highly detuned photonic modes in multimode cavities}},}\ }\href {https://doi.org/10.1063/5.0208379} {\bibfield  {journal} {\bibinfo  {journal} {J. Chem. Phys.}\ }\textbf {\bibinfo {volume} {160}},\ \bibinfo {pages} {204303} (\bibinfo {year} {2024})}\BibitemShut {NoStop}%
\bibitem [{\citenamefont {Thomas}\ \emph {et~al.}(2024)\citenamefont {Thomas}, \citenamefont {Tan}, \citenamefont {Kravets}, \citenamefont {Grigorenko},\ and\ \citenamefont {Barnes}}]{Thomas2024}%
  \BibitemOpen
  \bibfield  {author} {\bibinfo {author} {\bibfnamefont {P.~A.}\ \bibnamefont {Thomas}}, \bibinfo {author} {\bibfnamefont {W.~J.}\ \bibnamefont {Tan}}, \bibinfo {author} {\bibfnamefont {V.~G.}\ \bibnamefont {Kravets}}, \bibinfo {author} {\bibfnamefont {A.~N.}\ \bibnamefont {Grigorenko}},\ and\ \bibinfo {author} {\bibfnamefont {W.~L.}\ \bibnamefont {Barnes}},\ }\bibfield  {title} {\enquote {\bibinfo {title} {{Non-Polaritonic Effects in Cavity-Modified Photochemistry}},}\ }\href {https://doi.org/10.1002/ADMA.202309393} {\bibfield  {journal} {\bibinfo  {journal} {Adv. Mater.}\ }\textbf {\bibinfo {volume} {36}},\ \bibinfo {pages} {2309393} (\bibinfo {year} {2024})}\BibitemShut {NoStop}%
\bibitem [{\citenamefont {Michon}\ and\ \citenamefont {Simpkins}(2024)}]{Michon2024}%
  \BibitemOpen
  \bibfield  {author} {\bibinfo {author} {\bibfnamefont {M.~A.}\ \bibnamefont {Michon}}\ and\ \bibinfo {author} {\bibfnamefont {B.~S.}\ \bibnamefont {Simpkins}},\ }\bibfield  {title} {\enquote {\bibinfo {title} {{Impact of Cavity Length Non-uniformity on Reaction Rate Extraction in Strong Coupling Experiments}},}\ }\href {https://doi.org/10.1021/jacs.4c12269} {\bibfield  {journal} {\bibinfo  {journal} {J. Am. Chem. Soc.}\ }\textbf {\bibinfo {volume} {146}},\ \bibinfo {pages} {30596--30606} (\bibinfo {year} {2024})}\BibitemShut {NoStop}%
\bibitem [{\citenamefont {Galego}\ \emph {et~al.}(2019)\citenamefont {Galego}, \citenamefont {Climent}, \citenamefont {Garcia-Vidal},\ and\ \citenamefont {Feist}}]{Galego2019}%
  \BibitemOpen
  \bibfield  {author} {\bibinfo {author} {\bibfnamefont {J.}~\bibnamefont {Galego}}, \bibinfo {author} {\bibfnamefont {C.}~\bibnamefont {Climent}}, \bibinfo {author} {\bibfnamefont {F.~J.}\ \bibnamefont {Garcia-Vidal}},\ and\ \bibinfo {author} {\bibfnamefont {J.}~\bibnamefont {Feist}},\ }\bibfield  {title} {\enquote {\bibinfo {title} {{Cavity Casimir-Polder Forces and Their Effects in Ground-State Chemical Reactivity}},}\ }\href {https://doi.org/10.1103/PhysRevX.9.021057} {\bibfield  {journal} {\bibinfo  {journal} {Phys. Rev. X}\ }\textbf {\bibinfo {volume} {9}},\ \bibinfo {pages} {021057} (\bibinfo {year} {2019})}\BibitemShut {NoStop}%
\bibitem [{\citenamefont {Campos-Gonzalez-Angulo}, \citenamefont {Ribeiro},\ and\ \citenamefont {Yuen-Zhou}(2019)}]{Campos-Gonzalez-Angulo2019}%
  \BibitemOpen
  \bibfield  {author} {\bibinfo {author} {\bibfnamefont {J.~A.}\ \bibnamefont {Campos-Gonzalez-Angulo}}, \bibinfo {author} {\bibfnamefont {R.~F.}\ \bibnamefont {Ribeiro}},\ and\ \bibinfo {author} {\bibfnamefont {J.}~\bibnamefont {Yuen-Zhou}},\ }\bibfield  {title} {\enquote {\bibinfo {title} {{Resonant Catalysis of Thermally Activated Chemical Reactions with Vibrational Polaritons}},}\ }\href {https://doi.org/10.1038/s41467-019-12636-1} {\bibfield  {journal} {\bibinfo  {journal} {Nat. Commun.}\ }\textbf {\bibinfo {volume} {10}},\ \bibinfo {pages} {4685} (\bibinfo {year} {2019})}\BibitemShut {NoStop}%
\bibitem [{\citenamefont {Hoffmann}\ \emph {et~al.}(2020)\citenamefont {Hoffmann}, \citenamefont {Lacombe}, \citenamefont {Rubio},\ and\ \citenamefont {Maitra}}]{Hoffmann2020}%
  \BibitemOpen
  \bibfield  {author} {\bibinfo {author} {\bibfnamefont {N.~M.}\ \bibnamefont {Hoffmann}}, \bibinfo {author} {\bibfnamefont {L.}~\bibnamefont {Lacombe}}, \bibinfo {author} {\bibfnamefont {A.}~\bibnamefont {Rubio}},\ and\ \bibinfo {author} {\bibfnamefont {N.~T.}\ \bibnamefont {Maitra}},\ }\bibfield  {title} {\enquote {\bibinfo {title} {{Effect of Many Modes on Self-Polarization and Photochemical Suppression in Cavities}},}\ }\href {https://doi.org/10.1063/5.0012723} {\bibfield  {journal} {\bibinfo  {journal} {J. Chem. Phys.}\ }\textbf {\bibinfo {volume} {153}},\ \bibinfo {pages} {104103} (\bibinfo {year} {2020})}\BibitemShut {NoStop}%
\bibitem [{\citenamefont {Mandal}, \citenamefont {{Montillo Vega}},\ and\ \citenamefont {Huo}(2020)}]{Mandal2020Polarized}%
  \BibitemOpen
  \bibfield  {author} {\bibinfo {author} {\bibfnamefont {A.}~\bibnamefont {Mandal}}, \bibinfo {author} {\bibfnamefont {S.}~\bibnamefont {{Montillo Vega}}},\ and\ \bibinfo {author} {\bibfnamefont {P.}~\bibnamefont {Huo}},\ }\bibfield  {title} {\enquote {\bibinfo {title} {{Polarized Fock States and the Dynamical Casimir Effect in Molecular Cavity Quantum Electrodynamics}},}\ }\href {https://doi.org/10.1021/ACS.JPCLETT.0C02399} {\bibfield  {journal} {\bibinfo  {journal} {J. Phys. Chem. Lett.}\ }\textbf {\bibinfo {volume} {11}},\ \bibinfo {pages} {9215--9223} (\bibinfo {year} {2020})}\BibitemShut {NoStop}%
\bibitem [{\citenamefont {Fischer}\ and\ \citenamefont {Saalfrank}(2021)}]{Fischer2021}%
  \BibitemOpen
  \bibfield  {author} {\bibinfo {author} {\bibfnamefont {E.~W.}\ \bibnamefont {Fischer}}\ and\ \bibinfo {author} {\bibfnamefont {P.}~\bibnamefont {Saalfrank}},\ }\bibfield  {title} {\enquote {\bibinfo {title} {{Ground State Properties and Infrared Spectra of Anharmonic Vibrational Polaritons of Small Molecules in Cavities}},}\ }\href {https://doi.org/10.1063/5.0040853} {\bibfield  {journal} {\bibinfo  {journal} {J. Chem. Phys.}\ }\textbf {\bibinfo {volume} {154}},\ \bibinfo {pages} {104311} (\bibinfo {year} {2021})}\BibitemShut {NoStop}%
\bibitem [{\citenamefont {Yang}\ and\ \citenamefont {Cao}(2021)}]{YangCao2021}%
  \BibitemOpen
  \bibfield  {author} {\bibinfo {author} {\bibfnamefont {P.~Y.}\ \bibnamefont {Yang}}\ and\ \bibinfo {author} {\bibfnamefont {J.}~\bibnamefont {Cao}},\ }\bibfield  {title} {\enquote {\bibinfo {title} {{Quantum Effects in Chemical Reactions under Polaritonic Vibrational Strong Coupling}},}\ }\href {https://doi.org/10.1021/ACS.JPCLETT.1C02210/SUPPL_FILE/JZ1C02210_SI_001.PDF} {\bibfield  {journal} {\bibinfo  {journal} {J. Phys. Chem. Lett.}\ }\textbf {\bibinfo {volume} {12}},\ \bibinfo {pages} {9531--9538} (\bibinfo {year} {2021})}\BibitemShut {NoStop}%
\bibitem [{\citenamefont {Wang}\ \emph {et~al.}(2022)\citenamefont {Wang}, \citenamefont {Neuman}, \citenamefont {Yelin},\ and\ \citenamefont {Flick}}]{Wang2022JPCL}%
  \BibitemOpen
  \bibfield  {author} {\bibinfo {author} {\bibfnamefont {D.~S.}\ \bibnamefont {Wang}}, \bibinfo {author} {\bibfnamefont {T.}~\bibnamefont {Neuman}}, \bibinfo {author} {\bibfnamefont {S.~F.}\ \bibnamefont {Yelin}},\ and\ \bibinfo {author} {\bibfnamefont {J.}~\bibnamefont {Flick}},\ }\bibfield  {title} {\enquote {\bibinfo {title} {{Cavity-Modified Unimolecular Dissociation Reactions via Intramolecular Vibrational Energy Redistribution}},}\ }\href {https://doi.org/10.1021/acs.jpclett.2c00558} {\bibfield  {journal} {\bibinfo  {journal} {J. Phys. Chem. Lett}\ }\textbf {\bibinfo {volume} {13}},\ \bibinfo {pages} {3317--3324} (\bibinfo {year} {2022})}\BibitemShut {NoStop}%
\bibitem [{\citenamefont {Du}\ and\ \citenamefont {Yuen-Zhou}(2022)}]{DuDark2021}%
  \BibitemOpen
  \bibfield  {author} {\bibinfo {author} {\bibfnamefont {M.}~\bibnamefont {Du}}\ and\ \bibinfo {author} {\bibfnamefont {J.}~\bibnamefont {Yuen-Zhou}},\ }\bibfield  {title} {\enquote {\bibinfo {title} {{Catalysis by Dark States in Vibropolaritonic Chemistry}},}\ }\href {https://doi.org/10.1103/PHYSREVLETT.128.096001} {\bibfield  {journal} {\bibinfo  {journal} {Phys. Rev. Lett.}\ }\textbf {\bibinfo {volume} {128}},\ \bibinfo {pages} {096001} (\bibinfo {year} {2022})}\BibitemShut {NoStop}%
\bibitem [{\citenamefont {P{\'{e}}rez-S{\'{a}}nchez}\ \emph {et~al.}(2023)\citenamefont {P{\'{e}}rez-S{\'{a}}nchez}, \citenamefont {Koner}, \citenamefont {Stern},\ and\ \citenamefont {Yuen-Zhou}}]{Perez-Sanchez2023}%
  \BibitemOpen
  \bibfield  {author} {\bibinfo {author} {\bibfnamefont {J.~B.}\ \bibnamefont {P{\'{e}}rez-S{\'{a}}nchez}}, \bibinfo {author} {\bibfnamefont {A.}~\bibnamefont {Koner}}, \bibinfo {author} {\bibfnamefont {N.~P.}\ \bibnamefont {Stern}},\ and\ \bibinfo {author} {\bibfnamefont {J.}~\bibnamefont {Yuen-Zhou}},\ }\bibfield  {title} {\enquote {\bibinfo {title} {{Simulating molecular polaritons in the collective regime using few-molecule models}},}\ }\href {https://doi.org/10.1073/PNAS.2219223120} {\bibfield  {journal} {\bibinfo  {journal} {Proc. Natl. Acad. Sci.}\ }\textbf {\bibinfo {volume} {120}},\ \bibinfo {pages} {e2219223120} (\bibinfo {year} {2023})}\BibitemShut {NoStop}%
\bibitem [{\citenamefont {Weight}\ \emph {et~al.}(2024)\citenamefont {Weight}, \citenamefont {Weix}, \citenamefont {Tonzetich}, \citenamefont {Krauss},\ and\ \citenamefont {Huo}}]{Weight2024}%
  \BibitemOpen
  \bibfield  {author} {\bibinfo {author} {\bibfnamefont {B.~M.}\ \bibnamefont {Weight}}, \bibinfo {author} {\bibfnamefont {D.~J.}\ \bibnamefont {Weix}}, \bibinfo {author} {\bibfnamefont {Z.~J.}\ \bibnamefont {Tonzetich}}, \bibinfo {author} {\bibfnamefont {T.~D.}\ \bibnamefont {Krauss}},\ and\ \bibinfo {author} {\bibfnamefont {P.}~\bibnamefont {Huo}},\ }\bibfield  {title} {\enquote {\bibinfo {title} {{Cavity Quantum Electrodynamics Enables para- and ortho-Selective Electrophilic Bromination of Nitrobenzene}},}\ }\href {https://doi.org/10.1021/JACS.4C04045} {\bibfield  {journal} {\bibinfo  {journal} {J. Am. Chem. Soc.}\ }\textbf {\bibinfo {volume} {146}},\ \bibinfo {pages} {16184--16193} (\bibinfo {year} {2024})}\BibitemShut {NoStop}%
\bibitem [{\citenamefont {Aroeira}, \citenamefont {Kairys},\ and\ \citenamefont {Ribeiro}(2023)}]{Aroeira2023}%
  \BibitemOpen
  \bibfield  {author} {\bibinfo {author} {\bibfnamefont {G.~J.~R.}\ \bibnamefont {Aroeira}}, \bibinfo {author} {\bibfnamefont {K.~T.}\ \bibnamefont {Kairys}},\ and\ \bibinfo {author} {\bibfnamefont {R.~F.}\ \bibnamefont {Ribeiro}},\ }\bibfield  {title} {\enquote {\bibinfo {title} {{Theoretical Analysis of Exciton Wave Packet Dynamics in Polaritonic Wires}},}\ }\href {https://doi.org/10.1021/acs.jpclett.3c01082} {\bibfield  {journal} {\bibinfo  {journal} {J. Phys. Chem. Lett.}\ }\textbf {\bibinfo {volume} {14}},\ \bibinfo {pages} {5681--5691} (\bibinfo {year} {2023})}\BibitemShut {NoStop}%
\bibitem [{\citenamefont {Flick}\ \emph {et~al.}(2017)\citenamefont {Flick}, \citenamefont {Ruggenthaler}, \citenamefont {Appel},\ and\ \citenamefont {Rubio}}]{Flick2017}%
  \BibitemOpen
  \bibfield  {author} {\bibinfo {author} {\bibfnamefont {J.}~\bibnamefont {Flick}}, \bibinfo {author} {\bibfnamefont {M.}~\bibnamefont {Ruggenthaler}}, \bibinfo {author} {\bibfnamefont {H.}~\bibnamefont {Appel}},\ and\ \bibinfo {author} {\bibfnamefont {A.}~\bibnamefont {Rubio}},\ }\bibfield  {title} {\enquote {\bibinfo {title} {{Atoms and Molecules in Cavities, from Weak to Strong Coupling in Quantum-Electrodynamics (QED) Chemistry}},}\ }\href {https://doi.org/10.1073/pnas.1615509114} {\bibfield  {journal} {\bibinfo  {journal} {Proc. Natl. Acad. Sci.}\ }\textbf {\bibinfo {volume} {114}},\ \bibinfo {pages} {3026--3034} (\bibinfo {year} {2017})}\BibitemShut {NoStop}%
\bibitem [{\citenamefont {Haugland}\ \emph {et~al.}(2020)\citenamefont {Haugland}, \citenamefont {Ronca}, \citenamefont {Kj{\o}nstad}, \citenamefont {Rubio},\ and\ \citenamefont {Koch}}]{Haugland2020}%
  \BibitemOpen
  \bibfield  {author} {\bibinfo {author} {\bibfnamefont {T.~S.}\ \bibnamefont {Haugland}}, \bibinfo {author} {\bibfnamefont {E.}~\bibnamefont {Ronca}}, \bibinfo {author} {\bibfnamefont {E.~F.}\ \bibnamefont {Kj{\o}nstad}}, \bibinfo {author} {\bibfnamefont {A.}~\bibnamefont {Rubio}},\ and\ \bibinfo {author} {\bibfnamefont {H.}~\bibnamefont {Koch}},\ }\bibfield  {title} {\enquote {\bibinfo {title} {{Coupled Cluster Theory for Molecular Polaritons: Changing Ground and Excited States}},}\ }\href {https://doi.org/10.1103/PhysRevX.10.041043} {\bibfield  {journal} {\bibinfo  {journal} {Phys. Rev. X}\ }\textbf {\bibinfo {volume} {10}},\ \bibinfo {pages} {041043} (\bibinfo {year} {2020})}\BibitemShut {NoStop}%
\bibitem [{\citenamefont {Riso}\ \emph {et~al.}(2022)\citenamefont {Riso}, \citenamefont {Haugland}, \citenamefont {Ronca},\ and\ \citenamefont {Koch}}]{Riso2022}%
  \BibitemOpen
  \bibfield  {author} {\bibinfo {author} {\bibfnamefont {R.~R.}\ \bibnamefont {Riso}}, \bibinfo {author} {\bibfnamefont {T.~S.}\ \bibnamefont {Haugland}}, \bibinfo {author} {\bibfnamefont {E.}~\bibnamefont {Ronca}},\ and\ \bibinfo {author} {\bibfnamefont {H.}~\bibnamefont {Koch}},\ }\bibfield  {title} {\enquote {\bibinfo {title} {{Molecular Orbital Theory in Cavity QED Environments}},}\ }\href {https://doi.org/10.1038/s41467-022-29003-2} {\bibfield  {journal} {\bibinfo  {journal} {Nat. Commun.}\ }\textbf {\bibinfo {volume} {13}},\ \bibinfo {pages} {1368} (\bibinfo {year} {2022})}\BibitemShut {NoStop}%
\bibitem [{\citenamefont {Sch{\"{a}}fer}\ \emph {et~al.}(2022)\citenamefont {Sch{\"{a}}fer}, \citenamefont {Flick}, \citenamefont {Ronca}, \citenamefont {Narang},\ and\ \citenamefont {Rubio}}]{Schafer2021}%
  \BibitemOpen
  \bibfield  {author} {\bibinfo {author} {\bibfnamefont {C.}~\bibnamefont {Sch{\"{a}}fer}}, \bibinfo {author} {\bibfnamefont {J.}~\bibnamefont {Flick}}, \bibinfo {author} {\bibfnamefont {E.}~\bibnamefont {Ronca}}, \bibinfo {author} {\bibfnamefont {P.}~\bibnamefont {Narang}},\ and\ \bibinfo {author} {\bibfnamefont {A.}~\bibnamefont {Rubio}},\ }\bibfield  {title} {\enquote {\bibinfo {title} {{Shining Light on the Microscopic Resonant Mechanism Responsible for Cavity-Mediated Chemical Reactivity}},}\ }\href {https://doi.org/10.1038/s41467-022-35363-6} {\bibfield  {journal} {\bibinfo  {journal} {Nat. Commun.}\ }\textbf {\bibinfo {volume} {13}},\ \bibinfo {pages} {7817} (\bibinfo {year} {2022})}\BibitemShut {NoStop}%
\bibitem [{\citenamefont {Bonini}\ and\ \citenamefont {Flick}(2021)}]{Bonini2021}%
  \BibitemOpen
  \bibfield  {author} {\bibinfo {author} {\bibfnamefont {J.}~\bibnamefont {Bonini}}\ and\ \bibinfo {author} {\bibfnamefont {J.}~\bibnamefont {Flick}},\ }\bibfield  {title} {\enquote {\bibinfo {title} {{Ab Initio Linear-Response Approach to Vibro-polaritons in the Cavity Born-Oppenheimer Approximation}},}\ }\href {https://doi.org/10.1021/ACS.JCTC.1C01035} {\bibfield  {journal} {\bibinfo  {journal} {J. Chem. Theory Comput.}\ }\textbf {\bibinfo {volume} {18}},\ \bibinfo {pages} {2764--2773} (\bibinfo {year} {2021})}\BibitemShut {NoStop}%
\bibitem [{\citenamefont {Yang}\ \emph {et~al.}(2021)\citenamefont {Yang}, \citenamefont {Ou}, \citenamefont {Pei}, \citenamefont {Wang}, \citenamefont {Weng}, \citenamefont {Shuai}, \citenamefont {Mullen},\ and\ \citenamefont {Shao}}]{Yang2021}%
  \BibitemOpen
  \bibfield  {author} {\bibinfo {author} {\bibfnamefont {J.}~\bibnamefont {Yang}}, \bibinfo {author} {\bibfnamefont {Q.}~\bibnamefont {Ou}}, \bibinfo {author} {\bibfnamefont {Z.}~\bibnamefont {Pei}}, \bibinfo {author} {\bibfnamefont {H.}~\bibnamefont {Wang}}, \bibinfo {author} {\bibfnamefont {B.}~\bibnamefont {Weng}}, \bibinfo {author} {\bibfnamefont {Z.}~\bibnamefont {Shuai}}, \bibinfo {author} {\bibfnamefont {K.}~\bibnamefont {Mullen}},\ and\ \bibinfo {author} {\bibfnamefont {Y.}~\bibnamefont {Shao}},\ }\bibfield  {title} {\enquote {\bibinfo {title} {{Quantum-Electrodynamical Time-Dependent Density Functional Theory within Gaussian Atomic Basis}},}\ }\href {https://doi.org/10.1063/5.0057542} {\bibfield  {journal} {\bibinfo  {journal} {J. Chem. Phys.}\ }\textbf {\bibinfo {volume} {155}},\ \bibinfo {pages} {064107} (\bibinfo {year} {2021})}\BibitemShut {NoStop}%
\bibitem [{\citenamefont {McTague}\ and\ \citenamefont {Foley}(2022)}]{McTague2022}%
  \BibitemOpen
  \bibfield  {author} {\bibinfo {author} {\bibfnamefont {J.}~\bibnamefont {McTague}}\ and\ \bibinfo {author} {\bibfnamefont {J.~J.}\ \bibnamefont {Foley}},\ }\bibfield  {title} {\enquote {\bibinfo {title} {{Non-Hermitian cavity quantum electrodynamics-configuration interaction singles approach for polaritonic structure with ab initio molecular Hamiltonians}},}\ }\href {https://doi.org/10.1063/5.0091953} {\bibfield  {journal} {\bibinfo  {journal} {J. Chem. Phys.}\ }\textbf {\bibinfo {volume} {156}},\ \bibinfo {pages} {154103} (\bibinfo {year} {2022})}\BibitemShut {NoStop}%
\bibitem [{\citenamefont {Liebenthal}, \citenamefont {Vu},\ and\ \citenamefont {Deprince}(2022)}]{Liebenthal2022}%
  \BibitemOpen
  \bibfield  {author} {\bibinfo {author} {\bibfnamefont {M.~D.}\ \bibnamefont {Liebenthal}}, \bibinfo {author} {\bibfnamefont {N.}~\bibnamefont {Vu}},\ and\ \bibinfo {author} {\bibfnamefont {A.~E.}\ \bibnamefont {Deprince}},\ }\bibfield  {title} {\enquote {\bibinfo {title} {{Equation-of-motion cavity quantum electrodynamics coupled-cluster theory for electron attachment}},}\ }\href {https://doi.org/10.1063/5.0078795} {\bibfield  {journal} {\bibinfo  {journal} {J. Chem. Phys.}\ }\textbf {\bibinfo {volume} {156}},\ \bibinfo {pages} {54105} (\bibinfo {year} {2022})}\BibitemShut {NoStop}%
\bibitem [{\citenamefont {Luk}\ \emph {et~al.}(2017)\citenamefont {Luk}, \citenamefont {Feist}, \citenamefont {Toppari},\ and\ \citenamefont {Groenhof}}]{Luk2017}%
  \BibitemOpen
  \bibfield  {author} {\bibinfo {author} {\bibfnamefont {H.~L.}\ \bibnamefont {Luk}}, \bibinfo {author} {\bibfnamefont {J.}~\bibnamefont {Feist}}, \bibinfo {author} {\bibfnamefont {J.~J.}\ \bibnamefont {Toppari}},\ and\ \bibinfo {author} {\bibfnamefont {G.}~\bibnamefont {Groenhof}},\ }\bibfield  {title} {\enquote {\bibinfo {title} {{Multiscale Molecular Dynamics Simulations of Polaritonic Chemistry}},}\ }\href {https://doi.org/10.1021/acs.jctc.7b00388} {\bibfield  {journal} {\bibinfo  {journal} {J. Chem. Theory Comput.}\ }\textbf {\bibinfo {volume} {13}},\ \bibinfo {pages} {4324--4335} (\bibinfo {year} {2017})}\BibitemShut {NoStop}%
\bibitem [{\citenamefont {Li}, \citenamefont {Subotnik},\ and\ \citenamefont {Nitzan}(2020)}]{Li2020Water}%
  \BibitemOpen
  \bibfield  {author} {\bibinfo {author} {\bibfnamefont {T.~E.}\ \bibnamefont {Li}}, \bibinfo {author} {\bibfnamefont {J.~E.}\ \bibnamefont {Subotnik}},\ and\ \bibinfo {author} {\bibfnamefont {A.}~\bibnamefont {Nitzan}},\ }\bibfield  {title} {\enquote {\bibinfo {title} {{Cavity Molecular Dynamics Simulations of Liquid Water under Vibrational Ultrastrong Coupling}},}\ }\href {https://doi.org/10.1073/pnas.2009272117} {\bibfield  {journal} {\bibinfo  {journal} {Proc. Natl. Acad. Sci.}\ }\textbf {\bibinfo {volume} {117}},\ \bibinfo {pages} {18324--18331} (\bibinfo {year} {2020})}\BibitemShut {NoStop}%
\bibitem [{\citenamefont {G{\'{o}}mez}\ and\ \citenamefont {Vendrell}(2023)}]{Gomez2023}%
  \BibitemOpen
  \bibfield  {author} {\bibinfo {author} {\bibfnamefont {J.~A.}\ \bibnamefont {G{\'{o}}mez}}\ and\ \bibinfo {author} {\bibfnamefont {O.}~\bibnamefont {Vendrell}},\ }\bibfield  {title} {\enquote {\bibinfo {title} {{Vibrational Energy Redistribution and Polaritonic Fermi Resonances in the Strong Coupling Regime}},}\ }\href {https://doi.org/10.1021/ACS.JPCA.2C08608} {\bibfield  {journal} {\bibinfo  {journal} {J. Phys. Chem. A}\ }\textbf {\bibinfo {volume} {127}},\ \bibinfo {pages} {1598--1608} (\bibinfo {year} {2023})}\BibitemShut {NoStop}%
\bibitem [{\citenamefont {Lindoy}, \citenamefont {Mandal},\ and\ \citenamefont {Reichman}(2024)}]{Lindoy2024}%
  \BibitemOpen
  \bibfield  {author} {\bibinfo {author} {\bibfnamefont {L.~P.}\ \bibnamefont {Lindoy}}, \bibinfo {author} {\bibfnamefont {A.}~\bibnamefont {Mandal}},\ and\ \bibinfo {author} {\bibfnamefont {D.~R.}\ \bibnamefont {Reichman}},\ }\bibfield  {title} {\enquote {\bibinfo {title} {{Investigating the collective nature of cavity-modified chemical kinetics under vibrational strong coupling}},}\ }\href {https://doi.org/10.1515/NANOPH-2024-0026} {\bibfield  {journal} {\bibinfo  {journal} {Nanophotonics}\ }\textbf {\bibinfo {volume} {13}},\ \bibinfo {pages} {2617--2633} (\bibinfo {year} {2024})}\BibitemShut {NoStop}%
\bibitem [{\citenamefont {Hoffmann}\ \emph {et~al.}(2018)\citenamefont {Hoffmann}, \citenamefont {Appel}, \citenamefont {Rubio},\ and\ \citenamefont {Maitra}}]{Hoffmann2018}%
  \BibitemOpen
  \bibfield  {author} {\bibinfo {author} {\bibfnamefont {N.~M.}\ \bibnamefont {Hoffmann}}, \bibinfo {author} {\bibfnamefont {H.}~\bibnamefont {Appel}}, \bibinfo {author} {\bibfnamefont {A.}~\bibnamefont {Rubio}},\ and\ \bibinfo {author} {\bibfnamefont {N.~T.}\ \bibnamefont {Maitra}},\ }\bibfield  {title} {\enquote {\bibinfo {title} {{Light-matter Interactions via the Exact Factorization Approach}},}\ }\href {https://doi.org/10.1140/epjb/e2018-90177-6} {\bibfield  {journal} {\bibinfo  {journal} {Eur. Phys. J. B}\ }\textbf {\bibinfo {volume} {91}},\ \bibinfo {pages} {180} (\bibinfo {year} {2018})}\BibitemShut {NoStop}%
\bibitem [{\citenamefont {Rosenzweig}\ \emph {et~al.}(2022)\citenamefont {Rosenzweig}, \citenamefont {Hoffmann}, \citenamefont {Lacombe},\ and\ \citenamefont {Maitra}}]{Rosenzweig2022}%
  \BibitemOpen
  \bibfield  {author} {\bibinfo {author} {\bibfnamefont {B.}~\bibnamefont {Rosenzweig}}, \bibinfo {author} {\bibfnamefont {N.~M.}\ \bibnamefont {Hoffmann}}, \bibinfo {author} {\bibfnamefont {L.}~\bibnamefont {Lacombe}},\ and\ \bibinfo {author} {\bibfnamefont {N.~T.}\ \bibnamefont {Maitra}},\ }\bibfield  {title} {\enquote {\bibinfo {title} {{Analysis of the Classical Trajectory Treatment of Photon Dynamics for Polaritonic Phenomena}},}\ }\href {https://doi.org/10.1063/5.0079379} {\bibfield  {journal} {\bibinfo  {journal} {J. Chem. Phys.}\ }\textbf {\bibinfo {volume} {156}},\ \bibinfo {pages} {054101} (\bibinfo {year} {2022})}\BibitemShut {NoStop}%
\bibitem [{\citenamefont {{Sangiogo Gil}}, \citenamefont {Lauvergnat},\ and\ \citenamefont {Agostini}(2024)}]{SangiogoGil2024}%
  \BibitemOpen
  \bibfield  {author} {\bibinfo {author} {\bibfnamefont {E.}~\bibnamefont {{Sangiogo Gil}}}, \bibinfo {author} {\bibfnamefont {D.}~\bibnamefont {Lauvergnat}},\ and\ \bibinfo {author} {\bibfnamefont {F.}~\bibnamefont {Agostini}},\ }\bibfield  {title} {\enquote {\bibinfo {title} {{Exact factorization of the photon-electron-nuclear wavefunction: Formulation and coupled-trajectory dynamics}},}\ }\href {https://doi.org/10.1063/5.0224779} {\bibfield  {journal} {\bibinfo  {journal} {J. Chem. Phys.}\ }\textbf {\bibinfo {volume} {161}},\ \bibinfo {pages} {84112} (\bibinfo {year} {2024})}\BibitemShut {NoStop}%
\bibitem [{\citenamefont {Sukharev}, \citenamefont {Subotnik},\ and\ \citenamefont {Nitzan}(2023)}]{Sukharev2023}%
  \BibitemOpen
  \bibfield  {author} {\bibinfo {author} {\bibfnamefont {M.}~\bibnamefont {Sukharev}}, \bibinfo {author} {\bibfnamefont {J.}~\bibnamefont {Subotnik}},\ and\ \bibinfo {author} {\bibfnamefont {A.}~\bibnamefont {Nitzan}},\ }\bibfield  {title} {\enquote {\bibinfo {title} {{Dissociation slowdown by collective optical response under strong coupling conditions}},}\ }\href {https://doi.org/10.1063/5.0133972} {\bibfield  {journal} {\bibinfo  {journal} {J. Chem. Phys.}\ }\textbf {\bibinfo {volume} {158}},\ \bibinfo {pages} {084104} (\bibinfo {year} {2023})}\BibitemShut {NoStop}%
\bibitem [{\citenamefont {Sukharev}(2023)}]{Sukharev2023a}%
  \BibitemOpen
  \bibfield  {author} {\bibinfo {author} {\bibfnamefont {M.}~\bibnamefont {Sukharev}},\ }\bibfield  {title} {\enquote {\bibinfo {title} {{Efficient parallel strategy for molecular plasmonics – A numerical tool for integrating Maxwell-Schr{\"{o}}dinger equations in three dimensions}},}\ }\href {https://doi.org/10.1016/J.JCP.2023.111920} {\bibfield  {journal} {\bibinfo  {journal} {J. Comput. Phys.}\ }\textbf {\bibinfo {volume} {477}},\ \bibinfo {pages} {111920} (\bibinfo {year} {2023})}\BibitemShut {NoStop}%
\bibitem [{\citenamefont {Zhou}\ \emph {et~al.}(2024)\citenamefont {Zhou}, \citenamefont {Chen}, \citenamefont {Sukharev}, \citenamefont {Subotnik},\ and\ \citenamefont {Nitzan}}]{Zhou2024}%
  \BibitemOpen
  \bibfield  {author} {\bibinfo {author} {\bibfnamefont {Z.}~\bibnamefont {Zhou}}, \bibinfo {author} {\bibfnamefont {H.~T.}\ \bibnamefont {Chen}}, \bibinfo {author} {\bibfnamefont {M.}~\bibnamefont {Sukharev}}, \bibinfo {author} {\bibfnamefont {J.~E.}\ \bibnamefont {Subotnik}},\ and\ \bibinfo {author} {\bibfnamefont {A.}~\bibnamefont {Nitzan}},\ }\bibfield  {title} {\enquote {\bibinfo {title} {{Nature of polariton transport in a Fabry-Perot cavity}},}\ }\href {https://doi.org/10.1103/PHYSREVA.109.033717} {\bibfield  {journal} {\bibinfo  {journal} {Phys. Rev. A}\ }\textbf {\bibinfo {volume} {109}},\ \bibinfo {pages} {033717} (\bibinfo {year} {2024})}\BibitemShut {NoStop}%
\bibitem [{\citenamefont {Kuisma}\ \emph {et~al.}(2022)\citenamefont {Kuisma}, \citenamefont {Rousseaux}, \citenamefont {Czajkowski}, \citenamefont {Rossi}, \citenamefont {Shegai}, \citenamefont {Erhart},\ and\ \citenamefont {Antosiewicz}}]{Kuisma2022}%
  \BibitemOpen
  \bibfield  {author} {\bibinfo {author} {\bibfnamefont {M.}~\bibnamefont {Kuisma}}, \bibinfo {author} {\bibfnamefont {B.}~\bibnamefont {Rousseaux}}, \bibinfo {author} {\bibfnamefont {K.~M.}\ \bibnamefont {Czajkowski}}, \bibinfo {author} {\bibfnamefont {T.~P.}\ \bibnamefont {Rossi}}, \bibinfo {author} {\bibfnamefont {T.}~\bibnamefont {Shegai}}, \bibinfo {author} {\bibfnamefont {P.}~\bibnamefont {Erhart}},\ and\ \bibinfo {author} {\bibfnamefont {T.~J.}\ \bibnamefont {Antosiewicz}},\ }\bibfield  {title} {\enquote {\bibinfo {title} {{Ultrastrong Coupling of a Single Molecule to a Plasmonic Nanocavity: A First-Principles Study}},}\ }\href {https://doi.org/10.1021/ACSPHOTONICS.2C00066} {\bibfield  {journal} {\bibinfo  {journal} {ACS Photonics}\ }\textbf {\bibinfo {volume} {9}},\ \bibinfo {pages} {1065--1077} (\bibinfo {year} {2022})}\BibitemShut {NoStop}%
\bibitem [{\citenamefont {Groenhof}\ \emph {et~al.}(2019)\citenamefont {Groenhof}, \citenamefont {Climent}, \citenamefont {Feist}, \citenamefont {Morozov},\ and\ \citenamefont {Toppari}}]{Groenhof2019}%
  \BibitemOpen
  \bibfield  {author} {\bibinfo {author} {\bibfnamefont {G.}~\bibnamefont {Groenhof}}, \bibinfo {author} {\bibfnamefont {C.}~\bibnamefont {Climent}}, \bibinfo {author} {\bibfnamefont {J.}~\bibnamefont {Feist}}, \bibinfo {author} {\bibfnamefont {D.}~\bibnamefont {Morozov}},\ and\ \bibinfo {author} {\bibfnamefont {J.~J.}\ \bibnamefont {Toppari}},\ }\bibfield  {title} {\enquote {\bibinfo {title} {{Tracking Polariton Relaxation with Multiscale Molecular Dynamics Simulations}},}\ }\href {https://doi.org/10.1021/acs.jpclett.9b02192} {\bibfield  {journal} {\bibinfo  {journal} {J. Phys. Chem. Lett.}\ }\textbf {\bibinfo {volume} {10}},\ \bibinfo {pages} {5476--5483} (\bibinfo {year} {2019})}\BibitemShut {NoStop}%
\bibitem [{\citenamefont {Li}, \citenamefont {Nitzan},\ and\ \citenamefont {Subotnik}(2021)}]{Li2021Collective}%
  \BibitemOpen
  \bibfield  {author} {\bibinfo {author} {\bibfnamefont {T.~E.}\ \bibnamefont {Li}}, \bibinfo {author} {\bibfnamefont {A.}~\bibnamefont {Nitzan}},\ and\ \bibinfo {author} {\bibfnamefont {J.~E.}\ \bibnamefont {Subotnik}},\ }\bibfield  {title} {\enquote {\bibinfo {title} {{Collective Vibrational Strong Coupling Effects on Molecular Vibrational Relaxation and Energy Transfer: Numerical Insights via Cavity Molecular Dynamics Simulations**}},}\ }\href {https://doi.org/10.1002/anie.202103920} {\bibfield  {journal} {\bibinfo  {journal} {Angew. Chemie Int. Ed.}\ }\textbf {\bibinfo {volume} {60}},\ \bibinfo {pages} {15533--15540} (\bibinfo {year} {2021})}\BibitemShut {NoStop}%
\bibitem [{\citenamefont {Chng}\ \emph {et~al.}(2024)\citenamefont {Chng}, \citenamefont {Ying}, \citenamefont {Lai}, \citenamefont {Vamivakas}, \citenamefont {Cundiff}, \citenamefont {Krauss},\ and\ \citenamefont {Huo}}]{Chng2024JPCL}%
  \BibitemOpen
  \bibfield  {author} {\bibinfo {author} {\bibfnamefont {B.~X.~K.}\ \bibnamefont {Chng}}, \bibinfo {author} {\bibfnamefont {W.}~\bibnamefont {Ying}}, \bibinfo {author} {\bibfnamefont {Y.}~\bibnamefont {Lai}}, \bibinfo {author} {\bibfnamefont {A.~N.}\ \bibnamefont {Vamivakas}}, \bibinfo {author} {\bibfnamefont {S.~T.}\ \bibnamefont {Cundiff}}, \bibinfo {author} {\bibfnamefont {T.~D.}\ \bibnamefont {Krauss}},\ and\ \bibinfo {author} {\bibfnamefont {P.}~\bibnamefont {Huo}},\ }\bibfield  {title} {\enquote {\bibinfo {title} {{Mechanism of Molecular Polariton Decoherence in the Collective Light–Matter Couplings Regime}},}\ }\href {https://doi.org/10.1021/acs.jpclett.4c03049} {\bibfield  {journal} {\bibinfo  {journal} {J. Phys. Chem. Lett.}\ }\textbf {\bibinfo {volume} {15}},\ \bibinfo {pages} {11773--11783} (\bibinfo {year} {2024})}\BibitemShut {NoStop}%
\bibitem [{\citenamefont {Castin}\ and\ \citenamefont {Molmer}(1995)}]{Castin1995}%
  \BibitemOpen
  \bibfield  {author} {\bibinfo {author} {\bibfnamefont {Y.}~\bibnamefont {Castin}}\ and\ \bibinfo {author} {\bibfnamefont {K.}~\bibnamefont {Molmer}},\ }\bibfield  {title} {\enquote {\bibinfo {title} {{Maxwell--Bloch Equations: A Unified View of Nonlinear Optics and Nonlinear Atom Optics}},}\ }\href {https://doi.org/10.1103/PhysRevA.51.R3426} {\bibfield  {journal} {\bibinfo  {journal} {Phys. Rev. A}\ }\textbf {\bibinfo {volume} {51}},\ \bibinfo {pages} {R3426--R3428} (\bibinfo {year} {1995})}\BibitemShut {NoStop}%
\bibitem [{\citenamefont {Mukamel}(1999)}]{Mukamel1999}%
  \BibitemOpen
  \bibfield  {author} {\bibinfo {author} {\bibfnamefont {S.}~\bibnamefont {Mukamel}},\ }\href@noop {} {\emph {\bibinfo {title} {{Principles of Nonlinear Optical Spectroscopy}}}}\ (\bibinfo  {publisher} {Oxford University Press},\ \bibinfo {address} {New York},\ \bibinfo {year} {1999})\BibitemShut {NoStop}%
\bibitem [{\citenamefont {Lopata}\ and\ \citenamefont {Neuhauser}(2009{\natexlab{a}})}]{Lopata2009-1}%
  \BibitemOpen
  \bibfield  {author} {\bibinfo {author} {\bibfnamefont {K.}~\bibnamefont {Lopata}}\ and\ \bibinfo {author} {\bibfnamefont {D.}~\bibnamefont {Neuhauser}},\ }\bibfield  {title} {\enquote {\bibinfo {title} {{Multiscale Maxwell--Schr{\"{o}}dinger Modeling: A Split Field Finite--Difference Time--Domain Approach to Molecular Nanopolaritonics}},}\ }\href {https://doi.org/10.1063/1.3082245} {\bibfield  {journal} {\bibinfo  {journal} {J. Chem. Phys.}\ }\textbf {\bibinfo {volume} {130}},\ \bibinfo {pages} {104707} (\bibinfo {year} {2009}{\natexlab{a}})}\BibitemShut {NoStop}%
\bibitem [{\citenamefont {Lopata}\ and\ \citenamefont {Neuhauser}(2009{\natexlab{b}})}]{Lopata2009-2}%
  \BibitemOpen
  \bibfield  {author} {\bibinfo {author} {\bibfnamefont {K.}~\bibnamefont {Lopata}}\ and\ \bibinfo {author} {\bibfnamefont {D.}~\bibnamefont {Neuhauser}},\ }\bibfield  {title} {\enquote {\bibinfo {title} {{Nonlinear Nanopolaritonics: Finite--Difference Time--Domain Maxwell--Schr{\"{o}}dinger Simulation of Molecule-Assisted Plasmon Transfer}},}\ }\href {https://doi.org/10.1063/1.3167407} {\bibfield  {journal} {\bibinfo  {journal} {J. Chem. Phys.}\ }\textbf {\bibinfo {volume} {131}},\ \bibinfo {pages} {014701} (\bibinfo {year} {2009}{\natexlab{b}})}\BibitemShut {NoStop}%
\bibitem [{\citenamefont {Deinega}\ and\ \citenamefont {Seideman}(2014)}]{Deinega2014}%
  \BibitemOpen
  \bibfield  {author} {\bibinfo {author} {\bibfnamefont {A.}~\bibnamefont {Deinega}}\ and\ \bibinfo {author} {\bibfnamefont {T.}~\bibnamefont {Seideman}},\ }\bibfield  {title} {\enquote {\bibinfo {title} {{Self-Interaction-Free Approaches for Self-Consistent Solution of the Maxwell--Liouville Equations}},}\ }\href {https://doi.org/10.1103/PhysRevA.89.022501} {\bibfield  {journal} {\bibinfo  {journal} {Phys. Rev. A}\ }\textbf {\bibinfo {volume} {89}},\ \bibinfo {pages} {022501} (\bibinfo {year} {2014})}\BibitemShut {NoStop}%
\bibitem [{\citenamefont {Schelew}\ \emph {et~al.}(2017)\citenamefont {Schelew}, \citenamefont {Ge}, \citenamefont {Hughes}, \citenamefont {Pond},\ and\ \citenamefont {Young}}]{Schelew2017}%
  \BibitemOpen
  \bibfield  {author} {\bibinfo {author} {\bibfnamefont {E.}~\bibnamefont {Schelew}}, \bibinfo {author} {\bibfnamefont {R.-C.}\ \bibnamefont {Ge}}, \bibinfo {author} {\bibfnamefont {S.}~\bibnamefont {Hughes}}, \bibinfo {author} {\bibfnamefont {J.}~\bibnamefont {Pond}},\ and\ \bibinfo {author} {\bibfnamefont {J.~F.}\ \bibnamefont {Young}},\ }\bibfield  {title} {\enquote {\bibinfo {title} {{Self-Consistent Numerical Modeling of Radiatively Damped Lorentz Oscillators}},}\ }\href {https://doi.org/10.1103/PhysRevA.95.063853} {\bibfield  {journal} {\bibinfo  {journal} {Phys. Rev. A}\ }\textbf {\bibinfo {volume} {95}},\ \bibinfo {pages} {063853} (\bibinfo {year} {2017})}\BibitemShut {NoStop}%
\bibitem [{\citenamefont {Yamada}\ \emph {et~al.}(2018)\citenamefont {Yamada}, \citenamefont {Noda}, \citenamefont {Nobusada},\ and\ \citenamefont {Yabana}}]{Yamada2018}%
  \BibitemOpen
  \bibfield  {author} {\bibinfo {author} {\bibfnamefont {S.}~\bibnamefont {Yamada}}, \bibinfo {author} {\bibfnamefont {M.}~\bibnamefont {Noda}}, \bibinfo {author} {\bibfnamefont {K.}~\bibnamefont {Nobusada}},\ and\ \bibinfo {author} {\bibfnamefont {K.}~\bibnamefont {Yabana}},\ }\bibfield  {title} {\enquote {\bibinfo {title} {{Time-Dependent Density Functional Theory for Interaction of Ultrashort Light Pulse with thin Materials}},}\ }\href {https://doi.org/10.1103/PHYSREVB.98.245147} {\bibfield  {journal} {\bibinfo  {journal} {Phys. Rev. B}\ }\textbf {\bibinfo {volume} {98}},\ \bibinfo {pages} {245147} (\bibinfo {year} {2018})}\BibitemShut {NoStop}%
\bibitem [{\citenamefont {Li}\ \emph {et~al.}(2018{\natexlab{a}})\citenamefont {Li}, \citenamefont {Nitzan}, \citenamefont {Sukharev}, \citenamefont {Martinez}, \citenamefont {Chen},\ and\ \citenamefont {Subotnik}}]{Li2018Spontaneous}%
  \BibitemOpen
  \bibfield  {author} {\bibinfo {author} {\bibfnamefont {T.~E.}\ \bibnamefont {Li}}, \bibinfo {author} {\bibfnamefont {A.}~\bibnamefont {Nitzan}}, \bibinfo {author} {\bibfnamefont {M.}~\bibnamefont {Sukharev}}, \bibinfo {author} {\bibfnamefont {T.}~\bibnamefont {Martinez}}, \bibinfo {author} {\bibfnamefont {H.-T.}\ \bibnamefont {Chen}},\ and\ \bibinfo {author} {\bibfnamefont {J.~E.}\ \bibnamefont {Subotnik}},\ }\bibfield  {title} {\enquote {\bibinfo {title} {{Mixed Quantum--Classical Electrodynamics: Understanding Spontaneous Decay and Zero--Point Energy}},}\ }\href {https://doi.org/10.1103/PhysRevA.97.032105} {\bibfield  {journal} {\bibinfo  {journal} {Phys. Rev. A}\ }\textbf {\bibinfo {volume} {97}},\ \bibinfo {pages} {032105} (\bibinfo {year} {2018}{\natexlab{a}})}\BibitemShut {NoStop}%
\bibitem [{\citenamefont {Li}\ \emph {et~al.}(2018{\natexlab{b}})\citenamefont {Li}, \citenamefont {Chen}, \citenamefont {Nitzan}, \citenamefont {Sukharev},\ and\ \citenamefont {Subotnik}}]{Li2018Tradeoff}%
  \BibitemOpen
  \bibfield  {author} {\bibinfo {author} {\bibfnamefont {T.~E.}\ \bibnamefont {Li}}, \bibinfo {author} {\bibfnamefont {H.-T.}\ \bibnamefont {Chen}}, \bibinfo {author} {\bibfnamefont {A.}~\bibnamefont {Nitzan}}, \bibinfo {author} {\bibfnamefont {M.}~\bibnamefont {Sukharev}},\ and\ \bibinfo {author} {\bibfnamefont {J.~E.}\ \bibnamefont {Subotnik}},\ }\bibfield  {title} {\enquote {\bibinfo {title} {{A Necessary Trade-off for Semiclassical Electrodynamics: Accurate Short-Range Coulomb Interactions versus the Enforcement of Causality?}}}\ }\href {https://doi.org/10.1021/acs.jpclett.8b02309} {\bibfield  {journal} {\bibinfo  {journal} {J. Phys. Chem. Lett.}\ ,\ \bibinfo {pages} {5955--5961}} (\bibinfo {year} {2018}{\natexlab{b}})}\BibitemShut {NoStop}%
\bibitem [{\citenamefont {Sukharev}\ and\ \citenamefont {Nitzan}(2017)}]{Sukharev2017}%
  \BibitemOpen
  \bibfield  {author} {\bibinfo {author} {\bibfnamefont {M.}~\bibnamefont {Sukharev}}\ and\ \bibinfo {author} {\bibfnamefont {A.}~\bibnamefont {Nitzan}},\ }\bibfield  {title} {\enquote {\bibinfo {title} {{Optics of Exciton--Plasmon Nanomaterials}},}\ }\href {https://doi.org/10.1088/1361-648X/aa85ef} {\bibfield  {journal} {\bibinfo  {journal} {J. Phys. Condens. Matter}\ }\textbf {\bibinfo {volume} {29}},\ \bibinfo {pages} {443003} (\bibinfo {year} {2017})}\BibitemShut {NoStop}%
\bibitem [{\citenamefont {Taflove}\ and\ \citenamefont {Hagness}(2005)}]{Taflove2005}%
  \BibitemOpen
  \bibfield  {author} {\bibinfo {author} {\bibfnamefont {A.}~\bibnamefont {Taflove}}\ and\ \bibinfo {author} {\bibfnamefont {S.~C.}\ \bibnamefont {Hagness}},\ }\href@noop {} {\emph {\bibinfo {title} {{Computational Electrodynamics}}}},\ \bibinfo {edition} {3rd}\ ed.\ (\bibinfo  {publisher} {Artech House, Inc.},\ \bibinfo {address} {Norwood},\ \bibinfo {year} {2005})\BibitemShut {NoStop}%
\bibitem [{\citenamefont {Oskooi}\ \emph {et~al.}(2010)\citenamefont {Oskooi}, \citenamefont {Roundy}, \citenamefont {Ibanescu}, \citenamefont {Bermel}, \citenamefont {Joannopoulos},\ and\ \citenamefont {Johnson}}]{Oskooi2010}%
  \BibitemOpen
  \bibfield  {author} {\bibinfo {author} {\bibfnamefont {A.~F.}\ \bibnamefont {Oskooi}}, \bibinfo {author} {\bibfnamefont {D.}~\bibnamefont {Roundy}}, \bibinfo {author} {\bibfnamefont {M.}~\bibnamefont {Ibanescu}}, \bibinfo {author} {\bibfnamefont {P.}~\bibnamefont {Bermel}}, \bibinfo {author} {\bibfnamefont {J.}~\bibnamefont {Joannopoulos}},\ and\ \bibinfo {author} {\bibfnamefont {S.~G.}\ \bibnamefont {Johnson}},\ }\bibfield  {title} {\enquote {\bibinfo {title} {{Meep: A flexible free-software package for electromagnetic simulations by the FDTD method}},}\ }\href {https://doi.org/10.1016/j.cpc.2009.11.008} {\bibfield  {journal} {\bibinfo  {journal} {Comput. Phys. Commun.}\ }\textbf {\bibinfo {volume} {181}},\ \bibinfo {pages} {687--702} (\bibinfo {year} {2010})}\BibitemShut {NoStop}%
\bibitem [{\citenamefont {Tully}(1990)}]{Tully1990}%
  \BibitemOpen
  \bibfield  {author} {\bibinfo {author} {\bibfnamefont {J.~C.}\ \bibnamefont {Tully}},\ }\bibfield  {title} {\enquote {\bibinfo {title} {{Molecular Dynamics with Electronic Transitions}},}\ }\href {https://doi.org/10.1063/1.459170} {\bibfield  {journal} {\bibinfo  {journal} {J. Chem. Phys.}\ }\textbf {\bibinfo {volume} {93}},\ \bibinfo {pages} {1061--1071} (\bibinfo {year} {1990})}\BibitemShut {NoStop}%
\bibitem [{\citenamefont {Ben-Nun}, \citenamefont {Quenneville},\ and\ \citenamefont {Mart{\'{i}}nez}(2000)}]{Ben-Nun2000}%
  \BibitemOpen
  \bibfield  {author} {\bibinfo {author} {\bibfnamefont {M.}~\bibnamefont {Ben-Nun}}, \bibinfo {author} {\bibfnamefont {J.}~\bibnamefont {Quenneville}},\ and\ \bibinfo {author} {\bibfnamefont {T.~J.}\ \bibnamefont {Mart{\'{i}}nez}},\ }\bibfield  {title} {\enquote {\bibinfo {title} {{Ab Initio Multiple Spawning: Photochemistry from First Principles Quantum Molecular Dynamics}},}\ }\href {https://doi.org/10.1021/jp994174i} {\bibfield  {journal} {\bibinfo  {journal} {J. Phys. Chem. A}\ }\textbf {\bibinfo {volume} {104}},\ \bibinfo {pages} {5161--5175} (\bibinfo {year} {2000})}\BibitemShut {NoStop}%
\bibitem [{\citenamefont {Li}\ \emph {et~al.}(2005)\citenamefont {Li}, \citenamefont {Tully}, \citenamefont {Schlegel},\ and\ \citenamefont {Frisch}}]{Li2005Eh}%
  \BibitemOpen
  \bibfield  {author} {\bibinfo {author} {\bibfnamefont {X.}~\bibnamefont {Li}}, \bibinfo {author} {\bibfnamefont {J.~C.}\ \bibnamefont {Tully}}, \bibinfo {author} {\bibfnamefont {H.~B.}\ \bibnamefont {Schlegel}},\ and\ \bibinfo {author} {\bibfnamefont {M.~J.}\ \bibnamefont {Frisch}},\ }\bibfield  {title} {\enquote {\bibinfo {title} {{Ab initio Ehrenfest dynamics}},}\ }\href {https://doi.org/10.1063/1.2008258} {\bibfield  {journal} {\bibinfo  {journal} {J. Chem. Phys.}\ }\textbf {\bibinfo {volume} {123}},\ \bibinfo {pages} {084106} (\bibinfo {year} {2005})}\BibitemShut {NoStop}%
\bibitem [{\citenamefont {Skolnick}, \citenamefont {Fisher},\ and\ \citenamefont {Whittaker}(1998)}]{Skolnick1998}%
  \BibitemOpen
  \bibfield  {author} {\bibinfo {author} {\bibfnamefont {M.~S.}\ \bibnamefont {Skolnick}}, \bibinfo {author} {\bibfnamefont {T.~A.}\ \bibnamefont {Fisher}},\ and\ \bibinfo {author} {\bibfnamefont {D.~M.}\ \bibnamefont {Whittaker}},\ }\bibfield  {title} {\enquote {\bibinfo {title} {{Strong coupling phenomena in quantum microcavity structures}},}\ }\href {https://doi.org/10.1088/0268-1242/13/7/003} {\bibfield  {journal} {\bibinfo  {journal} {Semicond. Sci. Technol.}\ }\textbf {\bibinfo {volume} {13}},\ \bibinfo {pages} {645--669} (\bibinfo {year} {1998})}\BibitemShut {NoStop}%
\bibitem [{\citenamefont {Novotny}\ and\ \citenamefont {Hecht}(2006)}]{Novotny2006}%
  \BibitemOpen
  \bibfield  {author} {\bibinfo {author} {\bibfnamefont {L.}~\bibnamefont {Novotny}}\ and\ \bibinfo {author} {\bibfnamefont {B.}~\bibnamefont {Hecht}},\ }\href@noop {} {\emph {\bibinfo {title} {{Principles of Nano-Optics}}}}\ (\bibinfo  {publisher} {Cambridge University Press},\ \bibinfo {address} {Cambridge},\ \bibinfo {year} {2006})\BibitemShut {NoStop}%
\bibitem [{\citenamefont {Kim}\ \emph {et~al.}(2007)\citenamefont {Kim}, \citenamefont {Chrostowski}, \citenamefont {Bisaillon},\ and\ \citenamefont {Plant}}]{Kim2007}%
  \BibitemOpen
  \bibfield  {author} {\bibinfo {author} {\bibfnamefont {J.~H.~E.}\ \bibnamefont {Kim}}, \bibinfo {author} {\bibfnamefont {L.}~\bibnamefont {Chrostowski}}, \bibinfo {author} {\bibfnamefont {E.}~\bibnamefont {Bisaillon}},\ and\ \bibinfo {author} {\bibfnamefont {D.~V.}\ \bibnamefont {Plant}},\ }\bibfield  {title} {\enquote {\bibinfo {title} {{DBR, Sub-wavelength grating, and Photonic crystal slab Fabry-Perot cavity design using phase analysis by FDTD}},}\ }\href {https://doi.org/10.1364/OE.15.010330} {\bibfield  {journal} {\bibinfo  {journal} {Opt. Express}\ }\textbf {\bibinfo {volume} {15}},\ \bibinfo {pages} {10330} (\bibinfo {year} {2007})}\BibitemShut {NoStop}%
\bibitem [{\citenamefont {Cohen-Tannoudji}, \citenamefont {Dupont-Roc},\ and\ \citenamefont {Grynberg}(1997)}]{Cohen-Tannoudji1997}%
  \BibitemOpen
  \bibfield  {author} {\bibinfo {author} {\bibfnamefont {C.}~\bibnamefont {Cohen-Tannoudji}}, \bibinfo {author} {\bibfnamefont {J.}~\bibnamefont {Dupont-Roc}},\ and\ \bibinfo {author} {\bibfnamefont {G.}~\bibnamefont {Grynberg}},\ }\href@noop {} {\emph {\bibinfo {title} {{Photons and Atoms: Introduction to Quantum Electrodynamics}}}}\ (\bibinfo  {publisher} {Wiley},\ \bibinfo {address} {New York},\ \bibinfo {year} {1997})\ pp.\ \bibinfo {pages} {280--295}\BibitemShut {NoStop}%
\bibitem [{\citenamefont {Li}, \citenamefont {Chen},\ and\ \citenamefont {Subotnik}(2019)}]{Li2019Stimulated}%
  \BibitemOpen
  \bibfield  {author} {\bibinfo {author} {\bibfnamefont {T.~E.}\ \bibnamefont {Li}}, \bibinfo {author} {\bibfnamefont {H.-T.}\ \bibnamefont {Chen}},\ and\ \bibinfo {author} {\bibfnamefont {J.~E.}\ \bibnamefont {Subotnik}},\ }\bibfield  {title} {\enquote {\bibinfo {title} {{Comparison of Different Classical, Semiclassical, and Quantum Treatments of Light–Matter Interactions: Understanding Energy Conservation}},}\ }\href {https://doi.org/10.1021/acs.jctc.8b01232} {\bibfield  {journal} {\bibinfo  {journal} {J. Chem. Theory Comput.}\ }\textbf {\bibinfo {volume} {15}},\ \bibinfo {pages} {1957--1973} (\bibinfo {year} {2019})}\BibitemShut {NoStop}%
\bibitem [{\citenamefont {Jaynes}\ and\ \citenamefont {Cummings}(1963)}]{Jaynes1963}%
  \BibitemOpen
  \bibfield  {author} {\bibinfo {author} {\bibfnamefont {E.}~\bibnamefont {Jaynes}}\ and\ \bibinfo {author} {\bibfnamefont {F.}~\bibnamefont {Cummings}},\ }\bibfield  {title} {\enquote {\bibinfo {title} {{Comparison of Quantum and Semiclassical Radiation Theories with Application to the Beam Maser}},}\ }\href {https://doi.org/10.1109/PROC.1963.1664} {\bibfield  {journal} {\bibinfo  {journal} {Proc. IEEE}\ }\textbf {\bibinfo {volume} {51}},\ \bibinfo {pages} {89--109} (\bibinfo {year} {1963})}\BibitemShut {NoStop}%
\bibitem [{\citenamefont {Purcell}(1995)}]{Purcell1995}%
  \BibitemOpen
  \bibfield  {author} {\bibinfo {author} {\bibfnamefont {E.~M.}\ \bibnamefont {Purcell}},\ }\bibfield  {title} {\enquote {\bibinfo {title} {{Spontaneous Emission Probabilities at Radio Frequencies}},}\ }in\ \href {https://doi.org/10.1007/978-1-4615-1963-8_40} {\emph {\bibinfo {booktitle} {Confined Electrons and Photons. NATO ASI Series, vol 340}}},\ \bibinfo {editor} {edited by\ \bibinfo {editor} {\bibfnamefont {E.}~\bibnamefont {Burstein}}\ and\ \bibinfo {editor} {\bibfnamefont {C.}~\bibnamefont {Weisbuch}}}\ (\bibinfo  {publisher} {Springer},\ \bibinfo {address} {Boston, MA},\ \bibinfo {year} {1995})\ pp.\ \bibinfo {pages} {839--839}\BibitemShut {NoStop}%
\bibitem [{\citenamefont {Berezin}\ and\ \citenamefont {Achilefu}(2010)}]{Berezin2010}%
  \BibitemOpen
  \bibfield  {author} {\bibinfo {author} {\bibfnamefont {M.~Y.}\ \bibnamefont {Berezin}}\ and\ \bibinfo {author} {\bibfnamefont {S.}~\bibnamefont {Achilefu}},\ }\bibfield  {title} {\enquote {\bibinfo {title} {{Fluorescence Lifetime Measurements and Biological Imaging}},}\ }\href {https://doi.org/10.1021/cr900343z} {\bibfield  {journal} {\bibinfo  {journal} {Chem. Rev.}\ }\textbf {\bibinfo {volume} {110}},\ \bibinfo {pages} {2641--2684} (\bibinfo {year} {2010})}\BibitemShut {NoStop}%
\bibitem [{\citenamefont {Fano}(1961)}]{Fano1961}%
  \BibitemOpen
  \bibfield  {author} {\bibinfo {author} {\bibfnamefont {U.}~\bibnamefont {Fano}},\ }\bibfield  {title} {\enquote {\bibinfo {title} {{Effects of Configuration Interaction on Intensities and Phase Shifts}},}\ }\href {https://doi.org/10.1103/PhysRev.124.1866} {\bibfield  {journal} {\bibinfo  {journal} {Phys. Rev.}\ }\textbf {\bibinfo {volume} {124}},\ \bibinfo {pages} {1866--1878} (\bibinfo {year} {1961})}\BibitemShut {NoStop}%
\bibitem [{\citenamefont {Yamaguchi}, \citenamefont {Lyasota},\ and\ \citenamefont {Yuge}(2021)}]{Yamaguchi2021}%
  \BibitemOpen
  \bibfield  {author} {\bibinfo {author} {\bibfnamefont {M.}~\bibnamefont {Yamaguchi}}, \bibinfo {author} {\bibfnamefont {A.}~\bibnamefont {Lyasota}},\ and\ \bibinfo {author} {\bibfnamefont {T.}~\bibnamefont {Yuge}},\ }\bibfield  {title} {\enquote {\bibinfo {title} {{Theory of Fano effect in cavity quantum electrodynamics}},}\ }\href {https://doi.org/10.1103/PhysRevResearch.3.013037} {\bibfield  {journal} {\bibinfo  {journal} {Phys. Rev. Res.}\ }\textbf {\bibinfo {volume} {3}},\ \bibinfo {pages} {013037} (\bibinfo {year} {2021})}\BibitemShut {NoStop}%
\bibitem [{\citenamefont {Li}, \citenamefont {Nitzan},\ and\ \citenamefont {Subotnik}(2022)}]{Li2021Relaxation}%
  \BibitemOpen
  \bibfield  {author} {\bibinfo {author} {\bibfnamefont {T.~E.}\ \bibnamefont {Li}}, \bibinfo {author} {\bibfnamefont {A.}~\bibnamefont {Nitzan}},\ and\ \bibinfo {author} {\bibfnamefont {J.~E.}\ \bibnamefont {Subotnik}},\ }\bibfield  {title} {\enquote {\bibinfo {title} {{Polariton Relaxation under Vibrational Strong Coupling: Comparing Cavity Molecular Dynamics Simulations against Fermi's Golden Rule Rate}},}\ }\href {https://doi.org/10.1063/5.0079784} {\bibfield  {journal} {\bibinfo  {journal} {J. Chem. Phys.}\ }\textbf {\bibinfo {volume} {156}},\ \bibinfo {pages} {134106} (\bibinfo {year} {2022})}\BibitemShut {NoStop}%
\bibitem [{\citenamefont {Zeng}\ \emph {et~al.}(2023)\citenamefont {Zeng}, \citenamefont {P{\'{e}}rez-S{\'{a}}nchez}, \citenamefont {Eckdahl}, \citenamefont {Liu}, \citenamefont {Chang}, \citenamefont {Weiss}, \citenamefont {Kalow}, \citenamefont {Yuen-Zhou},\ and\ \citenamefont {Stern}}]{Zeng2023}%
  \BibitemOpen
  \bibfield  {author} {\bibinfo {author} {\bibfnamefont {H.}~\bibnamefont {Zeng}}, \bibinfo {author} {\bibfnamefont {J.~B.}\ \bibnamefont {P{\'{e}}rez-S{\'{a}}nchez}}, \bibinfo {author} {\bibfnamefont {C.~T.}\ \bibnamefont {Eckdahl}}, \bibinfo {author} {\bibfnamefont {P.}~\bibnamefont {Liu}}, \bibinfo {author} {\bibfnamefont {W.~J.}\ \bibnamefont {Chang}}, \bibinfo {author} {\bibfnamefont {E.~A.}\ \bibnamefont {Weiss}}, \bibinfo {author} {\bibfnamefont {J.~A.}\ \bibnamefont {Kalow}}, \bibinfo {author} {\bibfnamefont {J.}~\bibnamefont {Yuen-Zhou}},\ and\ \bibinfo {author} {\bibfnamefont {N.~P.}\ \bibnamefont {Stern}},\ }\bibfield  {title} {\enquote {\bibinfo {title} {{Control of Photoswitching Kinetics with Strong Light–Matter Coupling in a Cavity}},}\ }\href {https://doi.org/10.1021/jacs.3c04254} {\bibfield  {journal} {\bibinfo  {journal} {J. Am. Chem. Soc.}\ }\textbf {\bibinfo {volume} {145}},\ \bibinfo {pages} {19655--19661} (\bibinfo {year} {2023})}\BibitemShut {NoStop}%
\bibitem [{\citenamefont {Nitzan}(2006)}]{Nitzan2006}%
  \BibitemOpen
  \bibfield  {author} {\bibinfo {author} {\bibfnamefont {A.}~\bibnamefont {Nitzan}},\ }\href@noop {} {\emph {\bibinfo {title} {{Chemical Dynamics in Condensed Phases: Relaxation, Transfer and Reactions in Condensed Molecular Systems}}}}\ (\bibinfo  {publisher} {Oxford University Press},\ \bibinfo {address} {New York},\ \bibinfo {year} {2006})\BibitemShut {NoStop}%
\bibitem [{\citenamefont {Chen}\ \emph {et~al.}(2019)\citenamefont {Chen}, \citenamefont {Li}, \citenamefont {Sukharev}, \citenamefont {Nitzan},\ and\ \citenamefont {Subotnik}}]{Chen2018Spontaneous}%
  \BibitemOpen
  \bibfield  {author} {\bibinfo {author} {\bibfnamefont {H.-T.}\ \bibnamefont {Chen}}, \bibinfo {author} {\bibfnamefont {T.~E.}\ \bibnamefont {Li}}, \bibinfo {author} {\bibfnamefont {M.}~\bibnamefont {Sukharev}}, \bibinfo {author} {\bibfnamefont {A.}~\bibnamefont {Nitzan}},\ and\ \bibinfo {author} {\bibfnamefont {J.~E.}\ \bibnamefont {Subotnik}},\ }\bibfield  {title} {\enquote {\bibinfo {title} {{Ehrenfest+R dynamics. I. A Mixed Quantum–Classical Electrodynamics Simulation of Spontaneous Emission}},}\ }\href {https://doi.org/10.1063/1.5057365} {\bibfield  {journal} {\bibinfo  {journal} {J. Chem. Phys.}\ }\textbf {\bibinfo {volume} {150}},\ \bibinfo {pages} {044102} (\bibinfo {year} {2019})}\BibitemShut {NoStop}%
\end{thebibliography}

%


    \end{document}